\documentclass[runningheads]{llncs}
\usepackage{graphicx}
\usepackage[utf8]{inputenc}
\usepackage[english]{babel}

\usepackage[symbol]{footmisc}
\usepackage{algorithm}
\usepackage{algorithmicx}
\usepackage{algpseudocode}
\usepackage{subcaption}

\usepackage{graphicx} 
\usepackage{pdfpages}
\usepackage{epsfig} 
\usepackage{amsmath} 
\usepackage{amssymb}  
\usepackage{xcolor}

\usepackage{mathtools}
\usepackage{url}
\usepackage[toc,page]{appendix}
\usepackage{pdfpages}
\usepackage{blindtext}
\usepackage{colortbl}
\usepackage{float}
\usepackage[caption=false,font=footnotesize]{subfig}
\usepackage{lipsum} 
\usepackage{booktabs}
\usepackage{tikz}
\usepackage{xcolor}
\usepackage{algorithmicx}

\begin{document}
\title{D4+: Emergent Adversarial Driving Maneuvers with Approximate Functional Optimization}
\author{
Diego Ortiz Barbosa\inst{1} \and
Luis Burbano\inst{1} \and 
Carlos Hernandez\inst{2} \and
Zengxiang Lei\inst{3} \and 
Younghee Park\inst{2} \and
Satish Ukkusuri \inst{3} \and
Alvaro A Cardenas\inst{1}
} 
\authorrunning{D. Ortiz Barbosa et al.}
\titlerunning{D4+}
\institute{
 University of California Santa Cruz, Santa Cruz CA 95064, USA \and
 San Jose State University, San Jose CA 95192, USA \and
 Purdue University, West Lafayette, IN 47907, USA
}
\maketitle           

\begin{abstract}
Intelligent mechanisms implemented in autonomous vehicles, such as proactive driving assist and collision alerts, reduce traffic accidents. However, verifying their correct functionality is difficult due to complex interactions with the environment. This problem is exacerbated in adversarial environments, where an attacker can control the environment surrounding autonomous vehicles to exploit vulnerabilities.

To preemptively identify vulnerabilities in these systems, in this paper, we implement a scenario-based framework with a formal method to identify the impact of malicious drivers interacting with autonomous vehicles. The formalization of the evaluation requirements utilizes metric temporal logic (MTL) to identify a safety condition that we want to test. Our goal is to find, through a rigorous testing approach, any trace that violates this MTL safety specification. Our results can help designers identify the range of safe operational behaviors that prevent malicious drivers from exploiting the autonomous features of modern vehicles.


\end{abstract}

\section{Introduction}


Human drivers use skills, experience, and cognitive abilities to avoid crashing in congested stop-and-go freeway driving. They constantly monitor their surroundings and learn to anticipate the actions of other drivers and traffic flow. Introducing autonomous features in vehicles promises increased safety and reduced car accidents. However, the widespread availability of autonomous vehicle technology also introduces vulnerabilities, and errors that on a large scale could lead to significant problems. For example, if an aggressive driver discovers that abrupt cutting in front of an autonomous vehicle can cause it to crash, malicious drivers could attempt to cause collisions among autonomous vehicles using the same tactic worldwide.

Performing a systematic test of the autonomous features of vehicles will help us identify vulnerabilities early on. To discover or anticipate vulnerabilities in uncommon situations, we have to test autonomous and semi-autonomous vehicles under many different scenarios. This large-scale testing is possible only through simulations.  In this paper, we focus on scenario testing to identify vulnerabilities that can be exploited by a malicious driver. In particular, we use a model-based data-driven system that can capture different potential adversarial scenarios that may exploit the system. In particular, our work is inspired by dynamic data-driven application systems (DDDAS)~\cite{darema2023handbook} by using high-fidelity digital twins to improve decisions and control of real-time systems. 


Our objective is to prevent attackers from abusing autonomous vehicles to cause harm in the physical world. We do so by first specifying the safety properties that physical systems need to keep and testing whether an adversary can drive the emergent system to a state that invalidates the specification. To search for these adversary maneuvers, we use digital twins and a search procedure inspired by the DDDAS framework. As with other DDDAS, our method combines data from multiple dynamic sources--results of digital twins for self-driving vehicles in our case--to adapt our model (e.g., Bayesian optimization) in search for potential driving maneuvers that will trigger a vulnerability in an autonomous vehicle.  In particular, we propose an optimization approach, where our goal is to maximize a function that represents the attacker's objective. As optimizing the objective function is difficult, we generate a model to guide the search for counterexamples. We then iteratively generate data from our dynamic simulation into the model to refine the search for adversarial behavior. This updated model feeds the simulator with behaviors that attempt to maximize the attacker's objective and are potentially dangerous to other vehicles in the simulation.

This paper is an extension of our preliminary work D4~\cite{hernandez2024d4} presented at the DDDAS 2024 conference. In our previous work, we had given the adversary fixed attack signatures, which prevented us from discovering more sophisticated attack trajectories. In this extension, we also give the attacker the ability to select arbitrary functions of acceleration and braking and find surprising new emergent malicious driving maneuvers that we did not find in our previous work. 

We make our code openly available on the following GitHub repository \url{https://github.com/lburbano/acc_verifai} and uploaded videos of our results to help illustrate the practical results of our framework.



\section{Related Work}\label{sec:rw}



Safety is an essential requirement for CPS, as we want to avoid harm to the plant and its users. Given a set of safety properties $\phi$ and a system $S$, which has a set of inputs $U$, we want to determine whether $S$ satisfies $\phi$ for inputs $U$. The intricate interactions between computing and physical systems makes it difficult to guarantee safety. Moreover, the addition of artificial intelligence (AI) and machine learning (ML) algorithms makes these problems more complicated. While there are works focusing on only ML/AI algorithms \cite{blasch2021certifiable}, we need to study the safety of closed-loop systems.


We identify at least two paths of work:  \textbf{verification} and \textbf{testing}. 
Next, we provide an overview of safety testing and verification. However, we refer the reader to several survey works addressing testing and verification~\cite{deshmukh2019formal,kapinski2014simulation} or testing of black-box systems~\cite{corso2021survey}.


\subsection{Verification}

Verification uses mathematical techniques to provide a formal guarantee that the system $S$ satisfies the properties $\phi$ for all possible inputs $u\in U$ and initial states $x_0\in X_0$. 

\noindent \textbf{Formal proofs:} Formal proofs provide mathematically rigorous arguments to show that the CPS satisfies the property $\phi$.  There are several techniques to provide these proofs, such as Lyapunov functions~\cite{Khalil:1173048,10.1145/2562059.2562139} or control barrier functions~\cite{10.1145/2562059.2562139,cbc87a20-8264-374b-b836-e5a5f6e270e4}. 
Developing proofs usually requires extensive human intervention. 

\noindent\textbf{Model checking:} Model checking stands as an approach to reduce the requirement of human intervention to obtain formal guarantees. Given a model of the system and a specification, the model checker returns a certificate of the correctness or a counterexample showing that the system does not satisfy the specification.

\noindent \textbf{Reachability:} Reachability analysis uses a mathematical model of the system to estimate the set of possible states in the future. Then, using mathematical techniques, reachability analysis predicts the CPS's possible future behavior, aiming to show that the CPS will not arrive at an unsafe state~\cite{ivanov2019verisig,tran2020nnv}, even when the system has an ML-based controller.

\noindent \textbf{Data driven or simulation guided verification:} Several of the previous techniques may benefit from using data or simulations to obtain formal guarantees. 
Kapinski et al. \cite{kapinski2014simulation} use simulations to obtain Lyapunov-like functions to certify stability or safety. 
C2E2 \cite{duggirala2015c2e2} uses reachability analysis and simulations to certify the properties of CPS. 
S. Paul et al. \cite{paul2023formal} use model checking for verification of Aerospace Systems. Moreover, the authors use the DDDAS paradigm to perform runtime verification. They incorporate data into the verification process to refine the models and proofs.


\subsection{Falsification}
While safety verification provides formal arguments to certify that a system satisfies a property $\phi$, these methods usually do not escalate well with the system size.

An alternative to verification is falsification, where we want to find conditions in which the system does not satisfy (or \textit{falsifies}) a property. Mathematically, given a system $S$, and a set of properties $\phi$, we want to find inputs $u$ and initial conditions $x_0$ under which $\phi$ is not true. These conditions and inputs are counterexamples sufficient to prove that a CPS does not satisfy $\phi$. Therefore, rather than proving that the system satisfies a property for all possible inputs and initial conditions, we have to find an example that shows the system does not satisfy $\phi$. While the search for a counterexample is still not easy, it may be easier than providing formal proof.

Falsification is different than testing. In falsification, we want to find a counterexample to show that the system does not satisfy a property. Commonly, we use optimization techniques to find these falsifying conditions~\cite{corso2021survey,kapinski2016simulation}. In testing, we select several inputs and initial conditions to determine (test) if the system satisfies the property in those scenarios.


\noindent \textbf{Global optimization:} In optimization-based testing, we introduce a mathematical function that relates to the safety property we want to test; if the objective function is negative, then we find a counterexample. We then try to minimize this function to find inputs and initial conditions that make the system violate the property $\phi$.

Several works have proposed different objective functions~\cite{koren2018adaptive}, but the most common function is based on temporal logic. With temporal logic, we can describe properties over the states and time. More specifically, several works use signal temporal logic (STL) or metric temporal logic (MTL). Fainekos et al. \cite{fainekos2009robustness} propose a metric called robustness that we can use to create an objective function for the optimization. 

While several methods for optimization use gradient descent algorithms, we cannot use these methods for optimizing the robustness function. Gradient-based optimization assumes the availability of a gradient, and the robustness is not convex or soft with respect to the system inputs, making the gradient not available. To solve this problem, several works use global optimization algorithms.

Global optimization algorithms do not rely on gradients. Instead, these methods perform optimization in three steps~\cite{kapinski2014simulation}: 1) we select an input and initial state to the system, 2) run a simulation and compute the cost function, 3) the optimization algorithm selects another set of inputs and states, and repeat the process. The particular way of selecting this new input changes on the optimization method, which includes simulation annealing~\cite{romeijn1994simulated}, Bayesian optimization~\cite{mockus1989bayesian}, and cross entropy~\cite{de2005tutorial}.

\subsection{Our work}

The previous subsections show the current work on safety verification and falsification. Safety testing attempts to find bugs that lead to the violation of specification $\phi$ that may appear during the CPS operation. In this work, however, we focus on security; rather than finding bugs that may appear in the CPS operation, we want to find the actions an attacker can perform to exploit the CPS and lead it to a violation of the specification. 

We identify the main differences with previous works: 1) the inputs $u$ to the system are controlled by the attacker, which maliciously drives the system to an unsafe state while the attacker is still safe; 
2) the necessity of two objectives: the malicious objective, where the attacker wants to make the system fail a specification, and the attacker safety objective, where the attacker wants to be safe while performing the attack.

This work is an evolution of our previous work, where we have attacked Adaptive cruise controllers (ACC). In our initial study, the attacker was limited to braking, and the objective was to determine the optimal timing and intensity of the braking action~\cite{salgado2022fuzzing}. Our hypothesis was that even such a simple maneuver could induce unsafe behavior. While we could find successful attacks, the maneuvers were too restrictive. Therefore, we later introduced D4~\cite{hernandez2024d4}, where the attacker could brake and accelerate. In that paper, however, the attacker had to follow a fixed pattern while attacking. In this paper, we introduce D4+, where we extend our previous work by letting the attacker find the pattern and the shape of the attack; the attacker freely selects the throttle or braking at each time instant.

\section{Preliminaries}



We use a model of the system $\mathcal{F}$ to simulate environmental conditions. For example, in autonomous vehicles, the behavior of the system includes the dynamics of the vehicles, and the environmental conditions include elements such as the presence of pedestrians and the behavior of other vehicles. 

The variety of scenarios can be parameterized by the vector $\pi$, which defines various elements in the simulation, such as initial conditions. Each selection of $\pi$ defines a different test case \cite{fremont2020formal}. 

The objective is to see if there are conditions $\pi$ such that the CPS violates the safety specification. Therefore, safety testing requires two things: a formal language to define safety for the and a method to find $\pi$ that produces a violation of the specification. There are several methods in the literature for both things \cite{corso2021survey}. In this paper, we use \emph{metric temporal logic} as the formal language and \emph{robustness} as the way to change the metric temporal logic formula as an optimization function that we can use when searching for $\pi$. We now give details of these two concepts.

We  use metric temporal logic to define the attacker's objective. We define metric temporal logic (MTL) inductively as \cite{clarke2018handbook}:
\begin{equation*}
    \varphi \coloneqq \top \,|\,  p \,|\, \neg\, \varphi \,|\ \varphi_1 \wedge \varphi_2 \,|\ \varphi_1 \,\mathcal{U}_I\, \varphi_2
\end{equation*}
where $p\in AP$ is an atomic proposition, $\varphi, \varphi_1, \varphi_2$ are MTL formulas,   $\neg, \wedge$ and $\top$ are the negation and conjunction operators, and true from propositional logic. $\mathcal{U}$ is the until operator. The operator $\varphi_1 \mathcal U\varphi_2$ states that $\varphi_1$ is true until $\varphi_2$ is true. For us, it is useful to define the operators eventually $F_I \varphi \equiv \top \mathcal{U}_I \varphi$, and always $G_I \varphi \equiv \neg F_I \neg \varphi$. Each time operator comes with an interval $I=[a,b]$ with $a,b\in \mathbb{R}^{\geq 0}$. We can also use the disjunction operator $\varphi_1\vee \varphi_2 \equiv \neg (\neg \varphi_1 \wedge \neg \varphi_2)$.


Before presenting the quantitative semantics of MTL, we need to define a trajectory $\mathcal{T}$, and the set of trajectories $T$. Let us define the state of the system as $x\in X$, where $X$ is the set of all possible states. A trajectory is given by $(t_0,x_0)$, $(t_1, x_1)$,...$(t_k, x_k)$ where $t_i\in \mathbb{R}^{\geq 0}$ and $x_i\in X$ for all $i\in\{0,...,k-1\}$, with $\mathbb{R}^{\geq 0}$ the set of nonnegative real numbers and $X$ the set of states of the vehicles, as we introduced before.


We use the quantitative semantics of MTL. The quantitative semantics have a function $\rho:T\times MTL \to \mathbb{R}$ that takes the trajectory $\mathcal T$, a specification $\psi$ in MTL, and produces a real value called \textbf{robustness}. If $\rho(\mathcal T, \psi)>0$, the trajectory satisfies the MTL formula. If $\rho(\mathcal T, \psi)<0$, the trajectory does not satisfy the formula. If $\rho(\mathcal T, \psi) = 0$, we do not know whether the trajectory satisfies the formula. Thus, if $\rho$ is greater than zero but close to zero, then the trajectory is close to not satisfying the MTL formula.
We refer the reader to previous works for the algorithm to recursively compute the robustness \cite{fainekos2006robustness}.

Although researchers have applied these methods for safety validation in several systems like autonomous vehicles, we need to apply them to find maneuvers that an attacker can perform to compromise system safety. We face two main problems when applying safety validation to find attacks. First, we need to define the safety specification. While in safety validation, we may have specifications such as the cars not crashing, in security, an attacker may have several objectives, such as producing a crash while not being involved. Second, we need to clearly define the capabilities the attacker has to create the attack. In this paper, we address those two problems and show how to apply safety validation algorithms in security to find adversarial maneuvers.

\subsection{Use-Case}

\begin{figure}[ht]
\vspace{-2ex}
    \centering
    \includegraphics[width=0.6\textwidth]{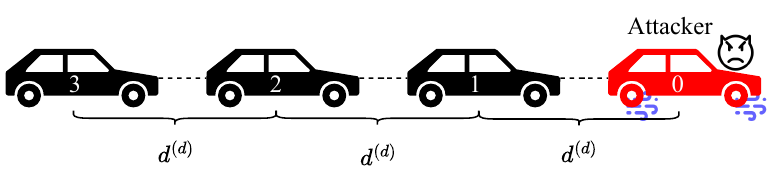}
    \caption{Experiment Setup of vehicles performing adaptive cruise control.
    }
    \label{fig:setup}
    \vspace{-5ex}
\end{figure}

To make the safety problems we consider in this paper concrete, let us consider a scenario with $N$ vehicles that move longitudinally, as we present in Fig. \ref{fig:setup}. As we only consider the longitudinal axis, we can describe each vehicle $i$ with the position and velocity of the center of mass in one axis, $x_i$ and $v_i$, respectively.

Each vehicle implements a switching controller that switches between a cruise control (when it is trying to maintain a given speed) and an adaptive cruise control (when it is trying to maintain a fixed distance to the vehicle in front) \cite{haspalamutgil2017adaptive,zhenhai2016multi}. Before presenting the switching policy, let us first explain both controllers.

\noindent \textbf{Cruise Control:} During the cruise control, the vehicle $i$ uses a discrete-time proportional-integral-derivative (PID) controller to achieve the desired velocity $v_i^{(d)}$. Thus, the acceleration during the cruise control is $a_i^{(cc)} \gets PID(v_i, v_i^{(d)})$. 

\noindent \textbf{Adaptive Cruise Control:} When the vehicle $i$ switches to adaptive cruise control (ACC), the vehicle wants to preserve a distance $d^{(d)}$ with respect to the vehicle in front and travel with the same velocity. Therefore, vehicle $i$ computes the acceleration as,
\begin{align*}
    a_i^{(acc)} = K^{(p)}_i\left(x_{i-1} - x_{i} - d^{(d)}\right) + K^{(v)}_i\left(v_{i-1} - v_{i}\right)
\end{align*}
where $K^{(p)}_i$ and $K^{(d)}_i$ are tuning parameters.

\noindent \textbf{Switching policy:} There are several policies in the literature for adaptive cruise control. We select a policy that uses position and velocity for the switching decision and hysteresis. Thus, the acceleration controller becomes:
\begin{equation}\label{eq:controller}
    a_i = \begin{cases}
        a^{(cc)}_i & \text{if } (x_{i-1}-x_{i}) \geq d^{(d)} + \epsilon^{(x)}\, \vee\, v_i \geq v^{(d)}_i + \epsilon^{(v)}\\
        a^{(acc)}_i & \text{if } (x_{i-1}-x_{i}) \leq d^{(d)} \,\wedge\, v_i \leq v^{(d)}_i
    \end{cases}
\end{equation}
where $\epsilon^{(x)}>0, \epsilon^{(v)}>0$ are hysteresis variables close to zero. We selected this policy because it avoids continuous switching between controllers, which may create issues. 

\noindent \textbf{Low-level controller:} The actual inputs $u(t)$ to the vehicle are throttle and braking instead of acceleration. Then, we implement a low-level controller that computes the throttle and braking the vehicle should apply at each time-instant, given the acceleration from Eq. \eqref{eq:controller}. We constrain this controller so that the vehicle cannot apply the throttle and brake at the same time.



\subsection{Threat Model}\label{sec:objective}

We assume the attacker can control one car. This means that the attacker can apply any arbitrary throttle, braking, or steering at any time instant. However, we assume the attacker cannot move backward. 



For simplicity, let us assume that the attacker is the vehicle in front of the cars' array. We will enumerate the attacker with the index $i=0$ and the vehicles in the rear in order from $i=1$ to $i=N$. Let us define the distance between the center of mass of vehicle $i$ and vehicle $j$ as:
$
    d_{i, j} \coloneqq |x_i-x_j|, 
$ where $x_i$ and $x_j$ are vehicle positions of vehicle $i$ and $j$, along the straight path.
Now, let us define the atomic proposition, $   \varphi_i \coloneqq (d_{i, i+1} > d_{safe}),$  where $d_{safe}$ is the safety distance between two consecutive vehicles.

We define the attacker specification in two parts: 1) the attacker wants the other vehicles to crash, and 2) the attacker does not want to be involved in a crash. We can define those two objectives with MTL. We encode the first part of the attacker objective as 
\begin{equation*}
    \varrho_{adv} = F \bigvee_{i=1}^{N-1} \neg \varphi_i.
\end{equation*}

In words, it means that at least two vehicles (other than the attacker) crash eventually. For the attacker's safety, we define the safety specification $\varrho_{safe}$ as, 
\begin{equation*}
    \varrho_{safe} = G \varphi_0     
\end{equation*}

\noindent meaning that the distance between the attacker's vehicle and vehicle $1$ is always larger than the safety distance. Then, the attacker wants to satisfy 
\begin{equation}\label{eq:obj}
    \overline \psi = \varrho_{adv} \wedge \varrho_{safe}.
\end{equation}


\section{Discovering Adversarial Maneuvers}

\begin{figure}[ht]
\vspace{-2ex}
    \centering
    \begin{subfigure}[b]{.6\linewidth}
    \includegraphics[width=\linewidth]{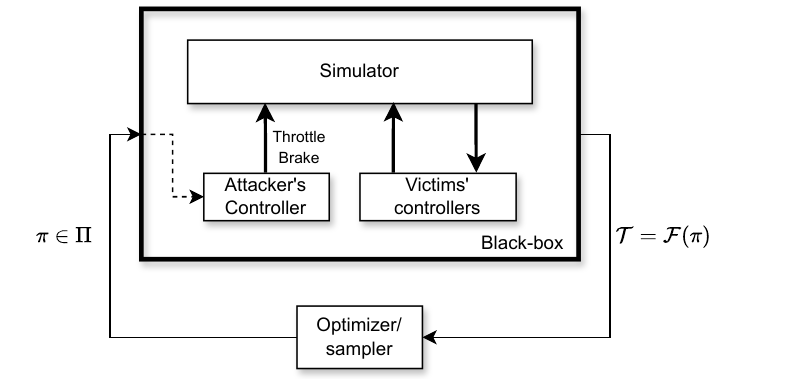}
    \caption{Parametric}\label{fig:time_space1}
\end{subfigure}
\begin{subfigure}[b]{.75\linewidth}
    \includegraphics[width=\linewidth]{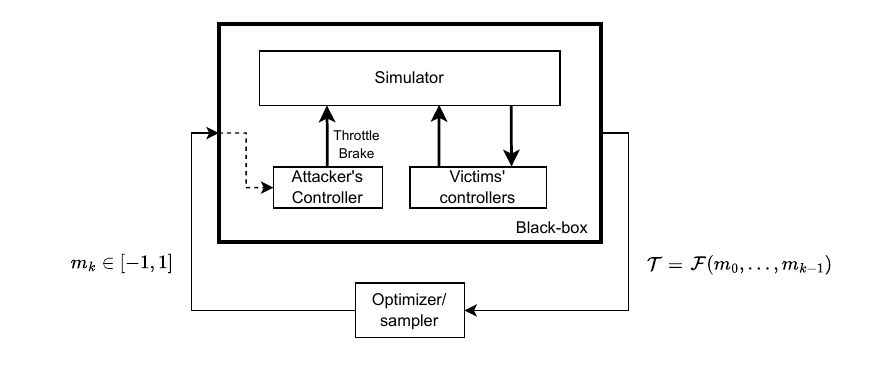}
    \caption{Nonparametric}\label{fig:time_space1}
\end{subfigure}
    \caption{Dynamic data-driven discovery (D4+) of adversarial maneuvers.}
    \label{fig:bb_attack_crafting}
    \vspace{-2ex}
\end{figure}

In this Section, we present how we can identify vulnerabilities to the attacks presented in the previous section. We summarize the approaches in Fig. \ref{fig:bb_attack_crafting}. We first present the method we use for finding adversarial maneuvers in our previous work, and then show the proposed modification.


For the discussion in the next subsections, let us denote the attacker's action at a time instant $k$ as $m_k\in {A}$, where ${A}$ is the set of possible actions that the attacker can apply. Therefore, as a consequence of the actions $m_0,...,m_{k-1}$, the simulator or environment creates a trajectory $T$. Thus, we can model this simulator as a function $F:{A}^{k}\to T$. That is, a function that maps from the attacker's actions to a trajectory. 

For our specific case, we consider that the set of possible actions is $A = [-1,1]$. This means that the attacker selects an action between $-1$ and $1$, where $-1$ is the maximum brake and $1$ is the maximum throttle. This way, we also impose the constraint that the attacker does not brake and accelerate simultaneously.

\subsection{Parametric attacks}

In our previous work, we propose parametric attacks. In this setup, the attacker followed a fixed-attack pattern, for example, a stop-and-go maneuver. The attacker generates an acceleration pattern using a function $s(\pi)$ parameters $\pi$ and generates the actions of $k-1$ actions. The attack parameters $\pi$ belong to a set of feasible parameters $\Pi$. For instance, the attacker can follow a sinusoidal signal, where the attack parameters are the frequency and the amplitude.  Formally, $s$ is a function $s: \Pi\to A^{k}$. As the attacker wants to satisfy the objective in Section \ref{sec:objective}, it wants to minimize the robustness $\rho$. Therefore, we proposed the optimization problem,
\begin{align}\label{eq:op}
    \begin{split}
        \max_{\pi} \quad  &\rho({F}(s(\pi)),\, \psi)\\
    s.t.\quad &\pi \in \Pi 
    \end{split}
\end{align}
As we mentioned in Sec. \ref{sec:rw}, we do not have the gradient when using the robustness function. Consequently, we use \textbf{global optimization} to find the attack signal.

Imposing the constraint to the attack, we hoped that the optimizers could find attacks more easily as the acceleration pattern $s(t,\pi)$ as the search space was smaller. However, this also limits the maneuvers that an attacker can perform. In the next section, we present a new version of our attack that lets the attacker search over a broader set of maneuvers while preserving the capacity to solve the optimization problem.

\subsection{Nonparametric attack}

In this section, we extend our previous work and let the attacker follow a broader acceleration pattern, creating a new challenge to create these new trajectories. Instead of selecting parameters to generate an attack, we directly generate the attacker's actions.  Therefore, the attacker solves the optimization problem,
\begin{align}\label{eq:op1}
    \begin{split}
        \max_{m_0,...,m_{k-1}} &\qquad \rho(F(m_0,...,m_{k-1}), \psi)\\
	s.t & \qquad m_0,...,m_{k-1}\in A,
    \end{split}
\end{align}
where $m_i$ is the action the attacker selects at each time step. Note that the attacker no longer optimizes over a set of parameters but over the action set itself. Therefore, we call this a \textbf{nonparametric} attack.

One option for this attack is to select a pair of acceleration and throttle actions for each time instant. While this approach could work, the optimization problem is not easy to solve, as we have several decision variables. 
To address the dimensionality problem, we use a similar approach to previous works in safety validation~\cite{akazaki2018falsification}. The attacker does not select an action at each time step. Instead, the attacker takes an action separated by $\Delta > 0$ seconds and uses interpolation to find the actions in between.

\section{Evaluation}

\begin{figure}[ht]
\vspace{-2ex}
    \centering
    \includegraphics[width=0.8\textwidth]{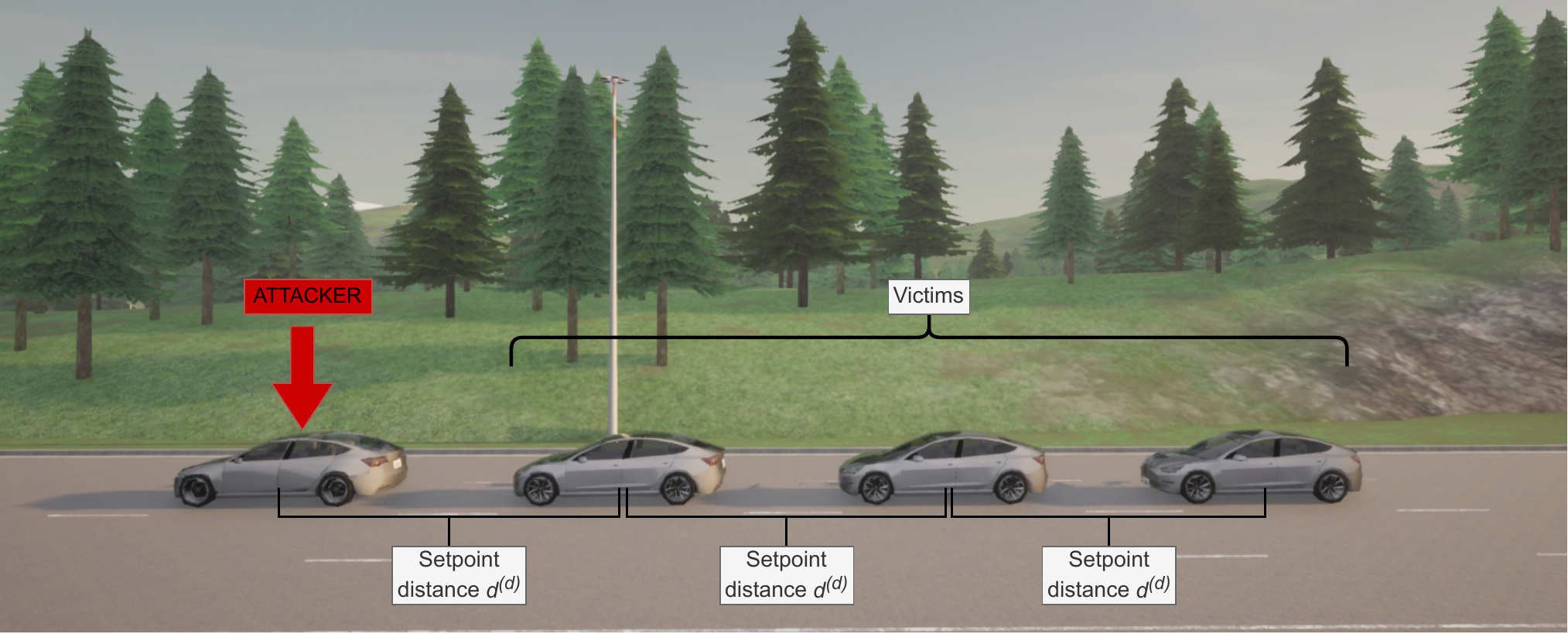}
    \caption{Attack setup in CARLA }
    \label{fig:carla}
    \vspace{-2ex}
\end{figure}
For our experiments, we use the CARLA simulator \cite{dosovitskiy2017carla} with our initial setup depicted in Fig. \ref{fig:carla}. CARLA  is an open-source vehicle simulation platform that is regularly maintained and can give accurate measures of the conditions of the  vehicles in a simulation, including precise distance measurements and positions along the road. 

We set up our testing scenario in CARLA's map \texttt{Town06} main road, a straight one-way 4-lane highway. We placed four vehicles in an array as shown in Fig. \ref{fig:carla} (a video of an attack is available at: \url{
https://youtu.be/3gtRjZhQ1x0?si=KGBciN4lmmf6x6eO}), with the leading vehicle being the attacker; we assume a busy highway where lane change is impossible due to traffic.  As previously stated, the attacker intends to cause its followers to crash without getting involved in the accident.


\begin{figure}[htb]
\vspace{-2ex}
\centering
\begin{subfigure}[b]{.35\linewidth}
    \includegraphics[width=\linewidth]{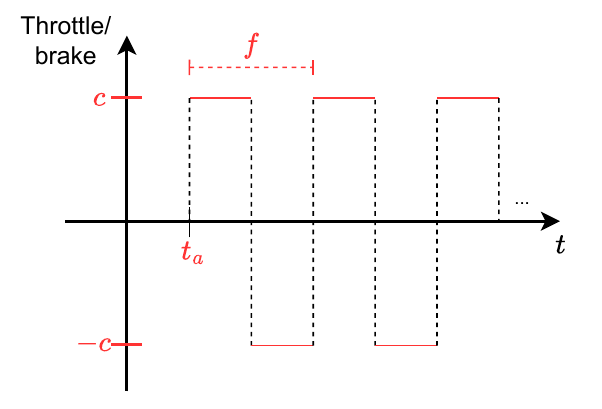}
    \caption{}\label{fig:square_attack}
    \end{subfigure}~~~~~~
    \begin{subfigure}[b]{.35\linewidth}
    \includegraphics[width=\linewidth]{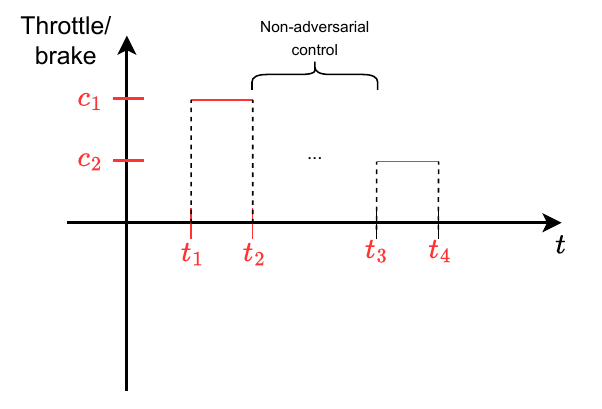}
    \caption{}\label{fig:attack2}
\end{subfigure}
\begin{subfigure}[b]{.35\linewidth}
    \includegraphics[width=\linewidth]{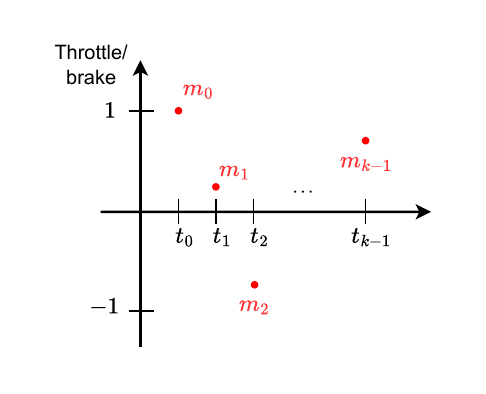}
    \caption{}\label{fig:attack3}
\end{subfigure}
\caption{Diagram of our three attack maneuvers: a) Persistent b) Intermittent c) Non-Parametric}
\vspace{-2ex}
\end{figure}

\subsection{Attack design}


\subsubsection{Parametrized attacks: }
For our parametrized attacks, we consider two attack signals: persistent (Fig.~\ref{fig:square_attack}) and intermittent (Fig.~\ref{fig:attack2}). The variables in red (in the figures) are the parameters the attack needs to define to generate an attack: for the persistent attack, $\pi_s=(c,f,t_a)$ and for the intermittent attack, $\pi_p = (t_1, t_2, t_3, t_4, c_1, c_2)$. 

We test these strategies in setpoint distances from 2 to 9 meters bumper to bumper, with a target speed of 20 m/s.  

\subsubsection{Nonparametrized attacks:}
We consider the following design for our non-parametric attack: (Fig.~\ref{fig:attack3}). The variables in red (in the figure) represent the throttle/brake that the attacker determines to generate the attack. As we said, we do not create a different attack signal at each time step. Instead, the attacker selects a new attack command every $\Delta t=6\,s$ and performs interpolation to obtain the intermediate values.

An attacker should follow a smooth behavior to avoid suspicious behavior. Therefore, we use spline interpolators as they preserve the softness of the attacker's commands. In particular, we select PCHIP (Piecewise Cubic Hermite Interpolated Polynomial), a spline that does not overshoot \cite{f1aac6cb-75ac-3b50-89ba-5fa72fe91ebe}, to ensure that the attacker applies actions that satisfy the objective function constraints. 

An attacker can deploy the attack in different stages of the system deployment. We consider the following two stages. 

\noindent\textit{Steady state attack:}
The adversary deploys the attack when the vehicles have arrived at the target velocity. 

\noindent\textit{Transient attack:} The adversary deploys the attack when all vehicles begin from rest (speed = 0). This scenario, for instance, emulates the vehicles using ACC after a light. 

While in our previous work, we only considered the \textit{steady state attack}, in this paper, we introduce the \textit{transient attack}.

To evaluate our new attack, we change the bumper-to-bumper distances from 3 to 15 meters. 
Additionally, We explored how different speeds could affect the attacker's behavior and effectiveness by conducting experiments with target speeds for the vehicle stream of 20, 25, and 30 m/s.

\subsection{Falsifier}

\noindent \textbf{VerifAI and Scenic: } VerifAI and Scenic are tools to conduct safety validation. Scenic is a probabilistic programming language. In Scenic, we can define all the conditions for creating the attack. In particular, we can define the set of attack parameters as a probabilistic distribution. 

In addition to Scenic, we use VerifAI. It is a scenario-testing safety validation tool fully compatible with CARLA. Additionally, it has global optimizers that, together with Scenic, allow us to solve the optimization problems from Eq.~\eqref{eq:op} and Eq.~\eqref{eq:op1}. 

\subsection{Optimizers}
To find these parameters, we need to solve the optimization problems from Eq. ~\ref{eq:op1} represented in Fig. ~\ref{fig:bb_attack_crafting}. We study use and study two optimization algorithms to solve these problems: \textbf{Bayesian optimization (BO) \cite{mockus1989bayesian} } and  \textbf{Cross entropy (CE) \cite{de2005tutorial}.} We next introduce these two optimization methods. For simplicity, we drop the parameters of the robustness function $\rho$. Moreover, we use the nonparametric attack to explain these methods. However, the only modification is the search space.

\noindent \textbf{Bayesian optimization (BO) \cite{mockus1989bayesian}:} creates a surrogate model (such as a probability distribution or a Gaussian process ) of the optimization function over the attack actions   $m_0,...,m_{k-1}$ in Eq.~\ref{eq:op1}. Initially, the surrogate model has an a priori belief of the optimization function. In the case of Equation~\ref{eq:op1}, Bayesian Optimization samples one element $m_0,...,m_{k-1}$ previously known. Afterward, we run the simulation to obtain a new trajectory ${F}(m_0,...,m_{k-1})$ and evaluate the objective function $\rho$. 
We update this surrogate model to obtain the a posteriori distribution; several strategies exist to make this update. We repeat this process for both problems until a stop criteria, is not such as a maximum number of iterations.

\noindent \textbf{Cross entropy (CE) \cite{de2005tutorial}:} is a global optimization based on importance sampling.
It assumes that the failure cases distribute according to distribution $q(\cdot, \theta)$, where $\theta$ parametrizes the distribution. The objective is then finding the optimal parameters $\theta^*$. For Eq.~\ref{eq:op1}, CE follows the next procedure. First, CE samples the attack actions $m_0,...,m_{k-1}$ from the distribution $q(\cdot, \theta)$ and runs simulations to obtain the trajectory ${F}(m_0,...,m_{k-1})$ and objective function $\rho$. Second, CE uses this value to estimate the optimal parameters $\theta^*$ by minimizing the Kullback-Leibler divergence between the distribution $q(\cdot, \theta)$ and $q(\cdot, \theta^*)$. For more details on how to perform that minimization, please refer to \cite{de2005tutorial}. Finally, CE repeats the process until some stopping criteria.




\section{Numerical results}

We begin by analyzing the number of successful attacks that we can find using our attacks. Table \ref{tab:res_statistics} shows a summary of all of the attacks in which the attacker successfully creates a crash between the other vehicles while avoiding crashing themselves. We find that persistent attacks tend to crash vehicles 1-2, while intermittent attacks tend to crash vehicles 2-3. Intermittent attacks found with CE tend to find a collision later in time, and the most dangerous crashes (those with the largest speed differences) are found with persistent attacks.

\begin{table}[!htbp]
\vspace{-2ex}

\caption{Attack results statistics}
\label{tab:res_statistics}
\resizebox{\columnwidth}{!}{%
\begin{tabular}{lrrrrr}
\hline
Attack name     & \begin{tabular}[r]{@{}r@{}}Number of \\  crashes\end{tabular} & \begin{tabular}[r]{@{}r@{}} Vehicle 1-2 \\ crashes\end{tabular} & \begin{tabular}[r]{@{}r@{}}Mean (Std) time \\ till collision (s)\end{tabular} & \begin{tabular}[r]{@{}r@{}}Collision location \\ along road (m)\end{tabular} & \begin{tabular}[r]{@{}r@{}}Speed differences in\\ crashes (km/h)\end{tabular} \\ \hline
Persistent BO   & 246                                                             & 155 (63\%)                                                                  & 18.3 (3.6)                                                                     & 87.6 (55.6)                                                                               & 8.9 (4.0)                                                                                       \\
Persistent CE   & 185                                                             & 112 (61\%)                                                                  & 17.8 (2.9)                                                                     & 79.0 (40.7)                                                                               & 8.7 (3.3)                                                                                       \\
Intermittent BO & 178                                                             & 82 (46\%)                                                                   & 17.1 (6.5)                                                                     & 70.6 (112.6)                                                                              & 6.7 (3.8)                                                                                       \\
Intermittent CE & 169                                                             & 83 (49\%)                                                                   & 28.9 (10.4)                                                                    & 258.8 (166.2)                                                                             & 5.9 (4.7)                                                                                       \\

\hline
\end{tabular}%
}

\end{table}

\begin{figure}[htb]
\vspace{-2ex}
\centering
\includegraphics[width=\linewidth]{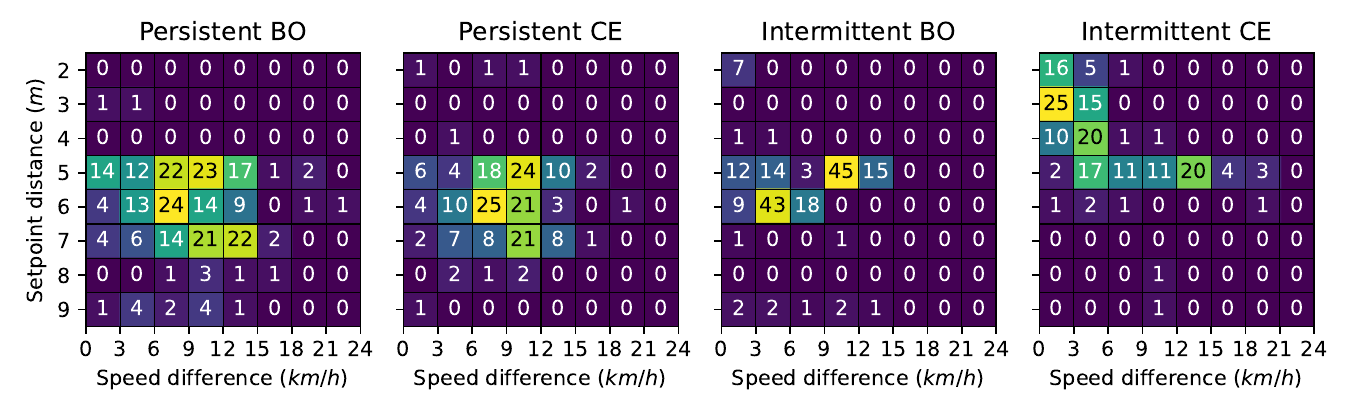}
\caption{Number of counter-examples (crashes) with different settings.}
\label{fig:res_mat}
\vspace{-2ex}
\end{figure}

\noindent \textbf{Most attacks happen when distance is 5-7 meters:}
Fig. \ref{fig:res_mat} shows that most attacks concentrate in the 5 to 7-meter setpoint range (the distance the ACC tries to keep from bumper to bumper with the vehicle ahead). Intuitively, the larger the setpoint distance, the fewer attacks occur. When the setpoint distance is 14 meters or higher, the controller can avoid any collision regardless of the attack's strategy, as it has enough time and distance to activate its mechanisms correctly.

\noindent \textbf{Attacker crashes when vehicles are close:} Intuitively, the smallest distance may lead to more crashes as vehicles have a shorter time to react to the attacker's maneuvers. However, when the vehicles are closer than $5\,m$, we cannot find several successful attacks. 

In distances in the 2 to 4-meter range, the attacker collides with the vehicle behind them. Table \ref{tab:res_attack_crash} shows how, in Intermittent and Persistent attacks, the attacker fails by crashing most of the time with the vehicle immediately behind them. Specifically, we can confirm that all the failures in setpoint 2 of the Persistent attack are due to the attacker crashing.

\begin{table}[!htbp]
\vspace{-2ex}

\caption{Attacker crashes in distances 2-5 meters}
\label{tab:res_attack_crash}
\resizebox{\columnwidth}{!}{%
\begin{tabular}{lrrrrrrrr}
\hline
Attack name     & \begin{tabular}[r]{@{}r@{}}Attacker\\   crash 2 m\end{tabular} & \begin{tabular}[r]{@{}r@{}} No crash \\ 2 m\end{tabular} & \begin{tabular}[r]{@{}r@{}}Attacker\\   crash 3 m\end{tabular} & \begin{tabular}[r]{@{}r@{}}No crash \\ 3 m\end{tabular} & \begin{tabular}[r]{@{}r@{}}Attacker\\   crash 4  m\end{tabular} & \begin{tabular}[r]{@{}r@{}}No crash\\   4  m\end{tabular} & \begin{tabular}[r]{@{}r@{}}Attacker\\  crash 5  m\end{tabular} & \begin{tabular}[r]{@{}r@{}}No crash\\   5  m\end{tabular}\\ \hline
Persistent BO   & 100                                                             & 0                                                                  & 100                                                                     
& 0                                                                               & 100 
    & 0
    & 9
    & 0
\\
Persistent CE  & 100                                                             & 0                                                                  & 99                                                                     
& 1                                                                               & 96 
    & 3
    & 32
    & 4                                                                                       \\
Intermittent BO & 84                                                             & 9                                                                  & 89                                                                     
& 11                                                                               & 90 
    & 8
    & 0
    & 11                                                                                       \\
Intermittent CE & 40                                                             & 38                                                                  & 33                                                                     
& 27                                                                               & 32 
    & 36
    & 0
    & 32                                                                                       \\

\hline
\end{tabular}%
}

\end{table}

Likewise, Fig. \ref{fig:res_mat} shows that BO in the Persistent attack produces the worst crash with a speed difference between vehicles of 25 km/h. In a practical sense, this speed difference implies that the crash is severe as the energy increases quadratically with the speed. 
Also, higher relative speed was reported to be associated with serious injury crashes following high-order power functions~\cite{elvik2009power}. 







\begin{figure}[htb]\vspace{-2ex}
\centering
\begin{subfigure}[b]{.325\linewidth}
    \includegraphics[width=\linewidth]{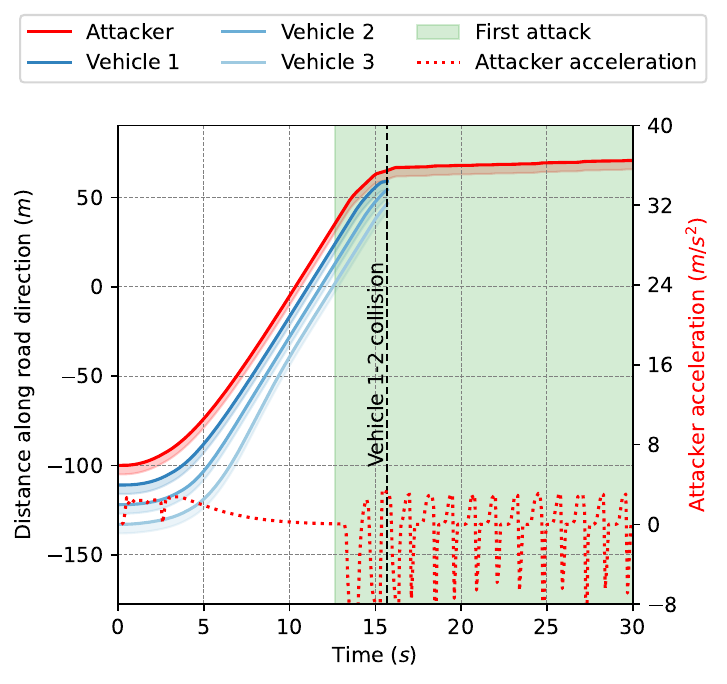}
    \caption{Persistent BO}\label{fig:time_space1}
\end{subfigure}
\begin{subfigure}[b]{.325\linewidth}
    \includegraphics[width=\linewidth]{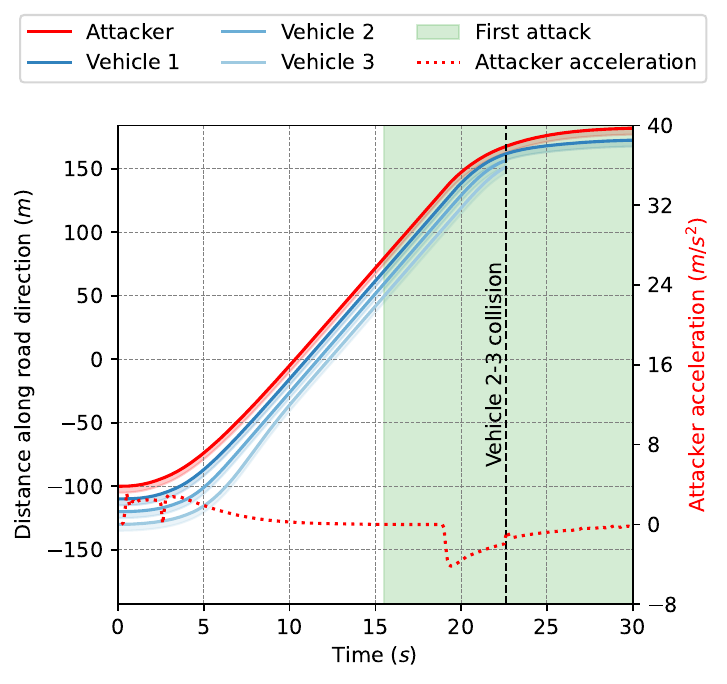}
    \caption{Persistent CE}\label{fig:time_space2}
\end{subfigure}
\begin{subfigure}[b]{.325\linewidth}
    \includegraphics[width=\linewidth]{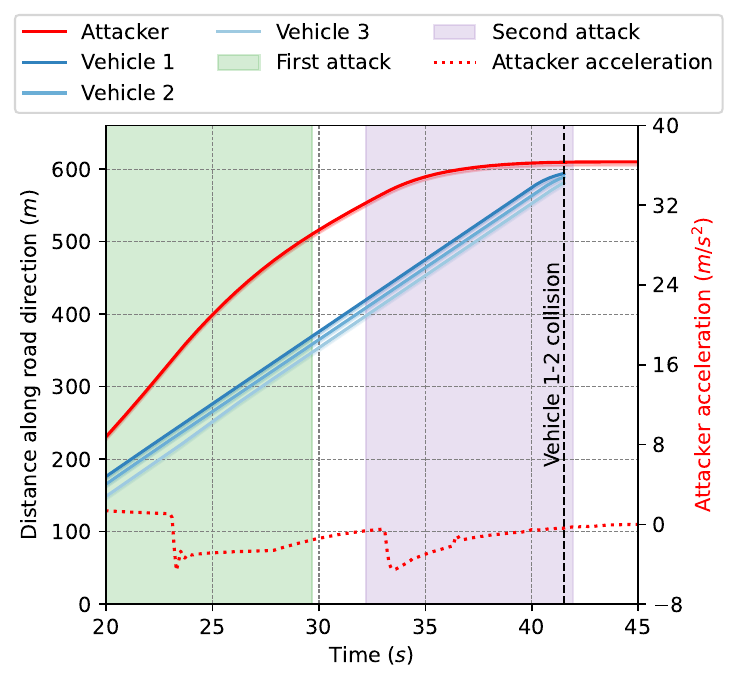}
    \caption{Intermittent CE }\label{fig:time_space4}
\end{subfigure}
\caption{Time-space diagram for large speed differences at collision ($> 15 km/h$).}
\label{fig:res_mat1}\vspace{-2ex}
\end{figure}

\begin{figure}
\vspace{-2ex}
 \centering
\includegraphics[width=0.55\textwidth]{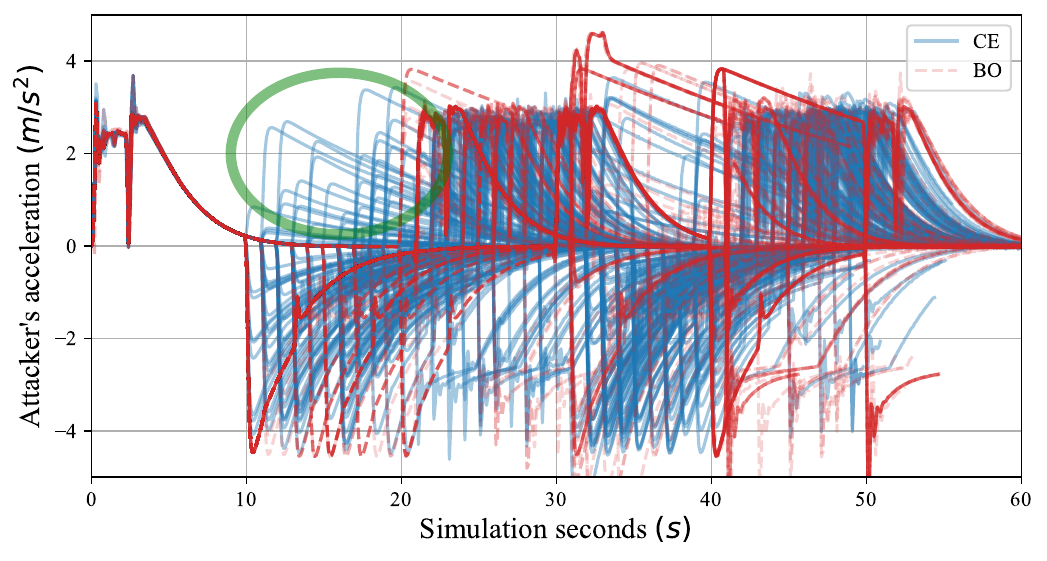}
\caption{Combined Vehicle Accelerations for Intermittent attack} 
\label{fig:combinedACC}
\vspace{-2ex}
\end{figure}

To better understand the generated vehicle trajectories, we employ a time-space diagram, a well-established tool in traffic operation analysis~\cite{daganzo1997fundamentals,wang2020data} to effectively provide information on vehicle position, speed, and acceleration. 
In Fig.~\ref{fig:res_mat1}, we can see some of the most dangerous attacks, with the most common attack strategy that consists of a sudden brake (Figs~\ref{fig:time_space1} and \ref{fig:time_space2}). However, in Fig.~\ref{fig:time_space4}, we can see a pattern that only happens in the Intermittent attack when being sampled by CE. The attacker starts in the first window, accelerating, trying to leave behind the vehicles behind them. In the second attack window, they make the sudden brake, generating the collision between vehicles 1 and 2. The circled area in Fig.~\ref{fig:combinedACC} shows more examples of this acceleration spike that CE gets in the first interval of the attack while comparing it to the different acceleration patterns created by BO.



\begin{table}[!htbp]
\small
\vspace{-2ex}

\caption{Free attack results statistics}
\label{tab:res_statistics_2}
\resizebox{\dimexpr\columnwidth-30pt}{!}{%
\begin{tabular}{lrrrr}
\hline
Attack Name     & \begin{tabular}[r]{@{}r@{}}Number of \\  crashes\end{tabular} & \begin{tabular}[r]{@{}r@{}} 20 (m/s)\\ crashes\end{tabular} & \begin{tabular}[r]{@{}r@{}} 25 (m/s)\\ crashes \end{tabular} & \begin{tabular}[r]{@{}r@{}} 30 (m/s)\\ crashes \end{tabular}\\ \hline

Interpol. BO Rest & 1594                                                             & 467 (29.30\%)                                                                   & 473 (29.67\%)                                                                     & 654 (41.03\%)                                                                                                                                                                    \\
Interpol. CE Rest & 369                                                             & 122 (33.06\%)                                                                   & 77 (28.87\%)                                                                    & 170 (46.07\%)                                                                                                                                                                  \\

Interpol. BO Stable & 1840                                                             & 664 (36.09\%)                                                                   & 619 (33.64\%)                                                                     & 557 (30.27\%)                                                                                                                                                                    \\
Interpol. CE Stable & 869                                                             & 244 (28.08\%)                                                                   & 281 (32.33\%)                                                                    & 344 (39.59\%)                                                                                                                                                                  \\
\hline
\end{tabular}%
} 

\end{table}

\section{Non-Parametric Attacks:} 
\begin{figure}[htb]\vspace{-2ex}
\centering
\begin{subfigure}[b]{.4\linewidth}
    \includegraphics[width=\linewidth]{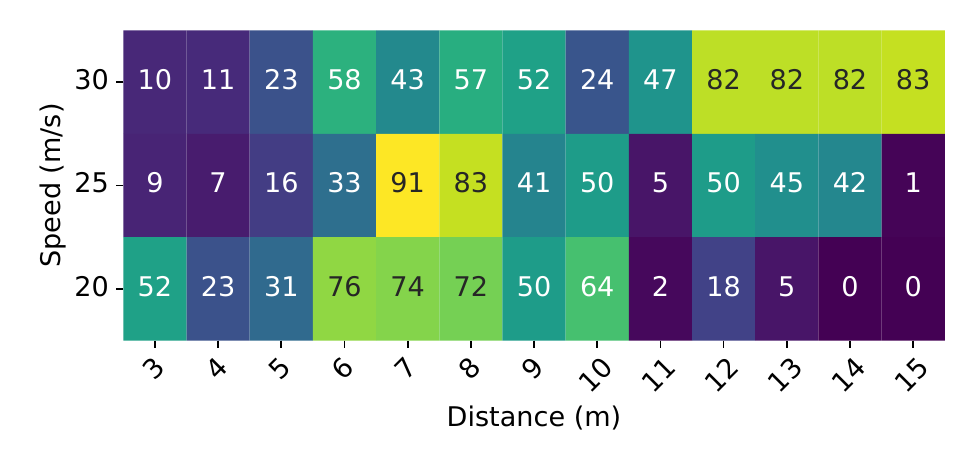}
    \caption{Transient BO}\label{fig:heatmap_1}
\end{subfigure}
\begin{subfigure}[b]{.4\linewidth}
    \includegraphics[width=\linewidth]{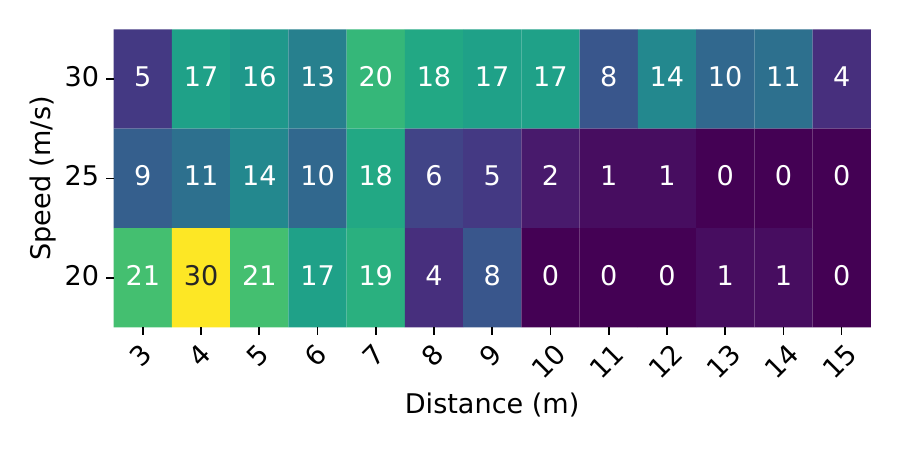}
    \caption{Transient CE}\label{fig:heatmap_2}
\end{subfigure}
\begin{subfigure}[b]{.4\linewidth}
    \includegraphics[width=\linewidth]{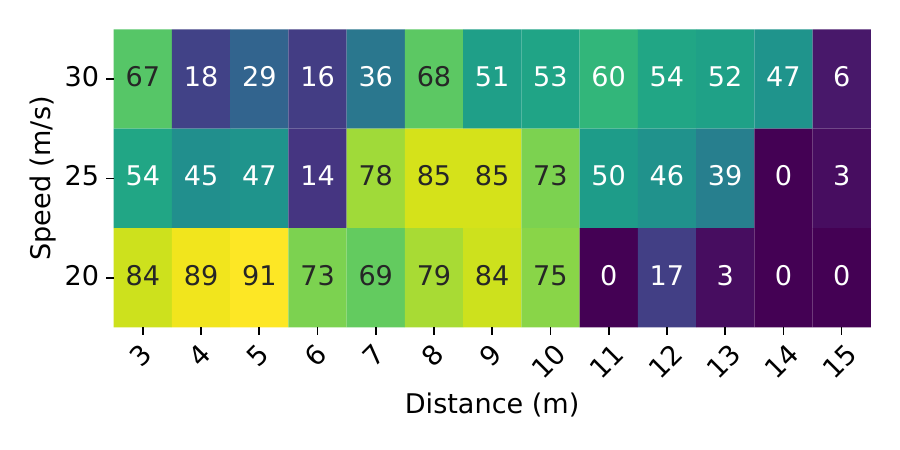}
    \caption{Steady BO}\label{fig:heatmap_3}
\end{subfigure}
\begin{subfigure}[b]{.4\linewidth}
    \includegraphics[width=\linewidth]{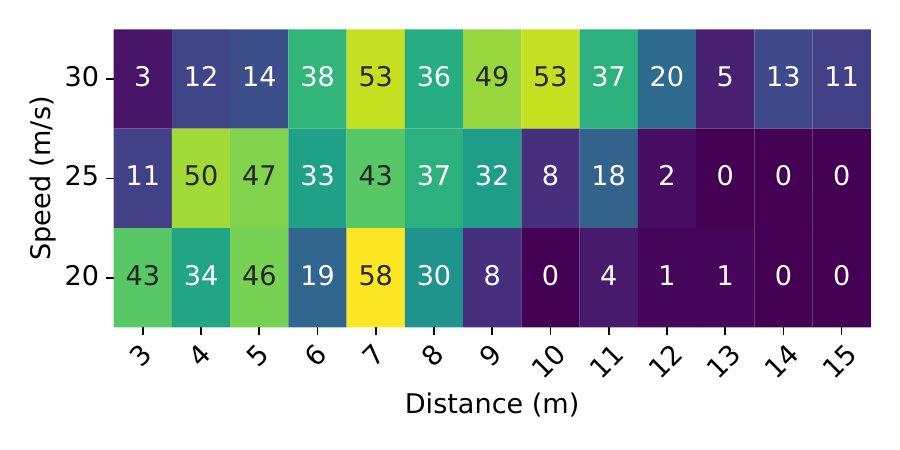}
    \caption{Steady CE}\label{fig:headmap_4}
\end{subfigure}

\caption{Heatmaps of collisions by experiments }
\label{fig:headmaps}\vspace{-2ex}
\end{figure}

We show basic statistics of the new maneuvers in Table \ref{tab:res_statistics_2}. Overall, BO shows a significantly higher number of successes than CE, finding 4.3 times more attacks when starting from the transient period and 2.1 more attacks after the steady state period. Table \ref{tab:res_statistics_2} also indicates that the traffic flow is more vulnerable when already in a steady state, we can infer this is due to the vehicles having to perform more complicated acceleration patterns to maintain the stable state.

Fig.~\ref{fig:headmaps} shows that successful attacks happen in most of the conditions tested, and the most significant number of attacks appear to occur when the vehicles want to keep a distance between each other of between 7 and 9 meters. Speed tends to influence most of the attacks in these ranges, whereas the minor target speeds significantly impact the number of attacks discovered. Intuitively, we can conclude that this range of distances allows the attacker to conduct its throttle and brake maneuvers in a way that avoids being involved in the crash. In contrast, the lack of success on more considerable distances can be attributed to the ACC having enough time to react at a slower speed, which diminishes the number of attacks further. 

Furthermore, Fig.~\ref{fig:headmaps} also shows how increased speed helps the attacker succeed even in greater distances, especially when using BO. This shows that although there are especially vulnerable setpoint distances for the ACC, the attack works for a distance that could, in principle, be considered safe to keep in an environment as tight as the one presented.

Another highlight is the difference in the number of attacks gathered by both sampling methods. While BO gathers 1594 successes, CE has a significant number of lower successes  with 369 successful attacks when starting from the transient period. We attribute this to the nature of each sampling technique: BO converges faster, focusing on maximizing the values that yield more results without exploring as much of the feature space. On the other hand, CE tends to explore more by sampling from the distribution more evenly and being slower to find successful attacks. With these insights we mainly focus the rest of this section on the results of BO.  

\begin{figure}[htb]\vspace{-2ex}
\centering
\begin{subfigure}[b]{.4\linewidth}
    \includegraphics[width=\linewidth]{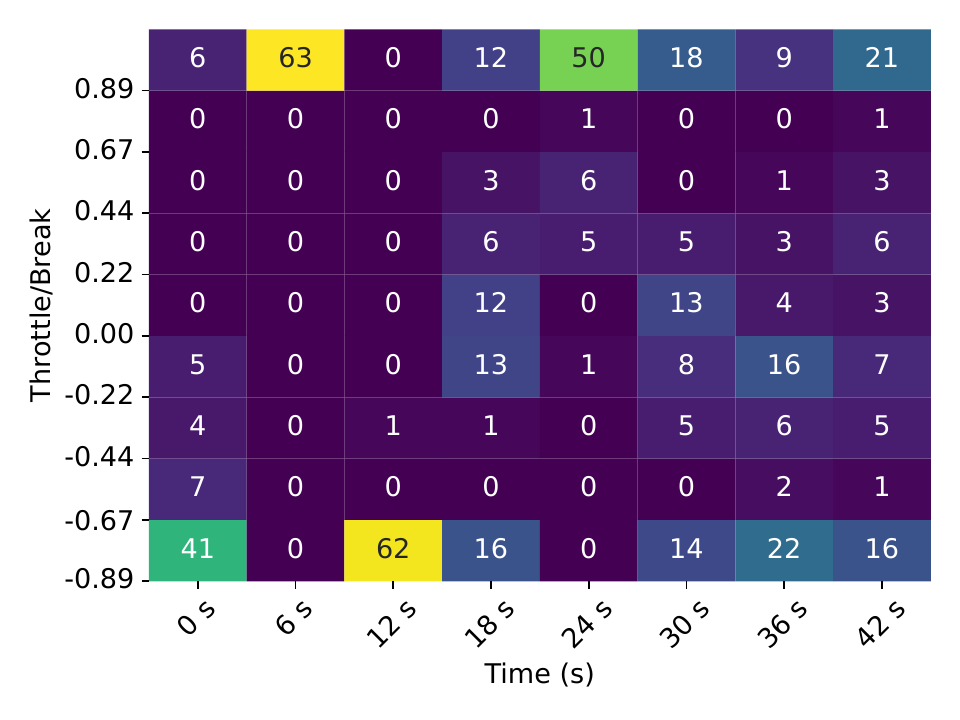}
    \caption{Cluster 1 Bayesian Optimization}\label{fig:clusterBO_1_rest}
\end{subfigure}
\begin{subfigure}[b]{.4\linewidth}
    \includegraphics[width=\linewidth]{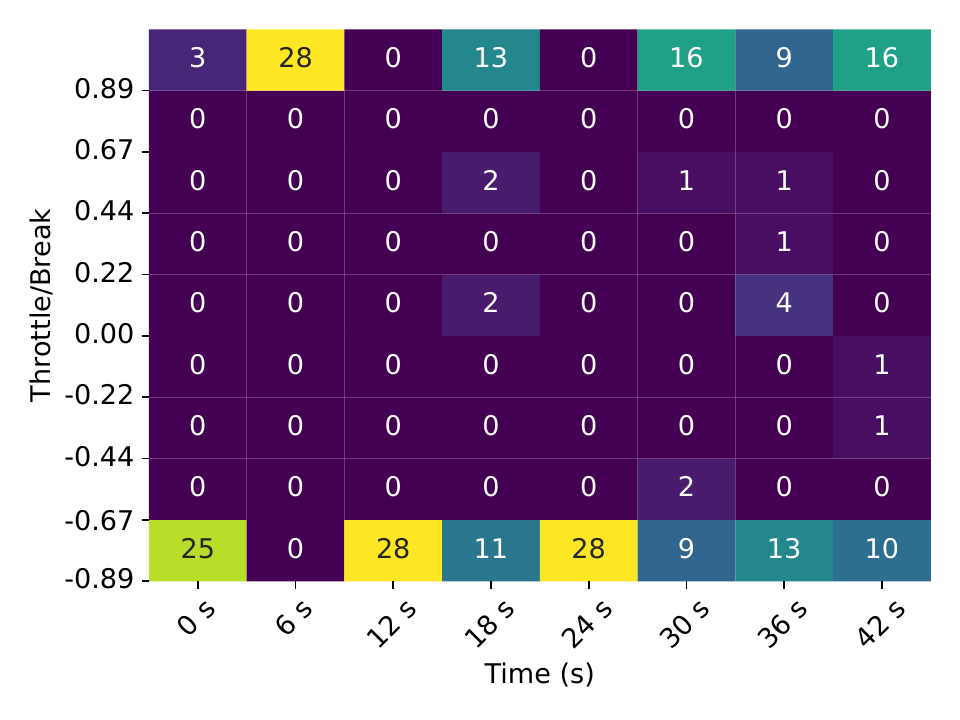}
    \caption{Cluster 2 Bayesian Optimization}\label{fig:clusterBO_2_rest}
\end{subfigure}

\caption{Values of parameters in each successful attack in each cluster for Bayesian Optimization with speed 25 and setpoint distance 7 (from rest) }

\label{fig:clusterBO_rest}\vspace{-2ex}
\end{figure}
\subsection{ Attacks during transient period:}

To understand how the attacker behaves, we analyze the parameters that generate the most significant number of successful attacks by sampler. In particular, we decided to perform a cluster analysis combining K-means \cite{1056489} and DBSCAN \cite{10.5555/3001460.3001507}, to find the patterns in the difference in the parameter values the attacker selects. Using the Elbow Rule to analyze the inertia as a function of the cluster, this alongside the DBSCAN result, helped us identify that we could divide our attack into two or, at most, three clusters, depending on the case. Fig.~\ref{fig:clusterBO_rest}  shows the two clusters produced for the most successful conditions using BO starting at rest.

\noindent \textbf{Attacker maneuver:} At the clusters in Fig.~\ref{fig:clusterBO_rest}, we can notice a consistent maneuver by the attacker that we depict in Alg. \ref{algo:man1}. By activating the attack before the vehicle is in a stable state, the attacker immediately halts and then engages full acceleration, creating the maximum distance between it and the previous vehicle. This allows the vehicles behind to accelerate freely and try to reach the target speed of 25 m/s. However, the attacker takes advantage of this situation and, in the next instance, decides to do another hard brake that causes a collision between the other vehicles. 

\begin{algorithm}
    \caption{Transient Period Maneuver}
    \begin{algorithmic}

        \State \textbf{Apply  Max Brake} at $t = 0$
        \State \textbf{Apply Full Throttle} at $t \approx  6$
        \State \textbf{Apply  Max Brake} at $t \approx  12$
    \end{algorithmic}

    \label{algo:man1}
\end{algorithm}
\begin{figure}[htb]\vspace{-2ex}
\centering
\begin{subfigure}[b]{.4\linewidth}
    \includegraphics[width=\linewidth]{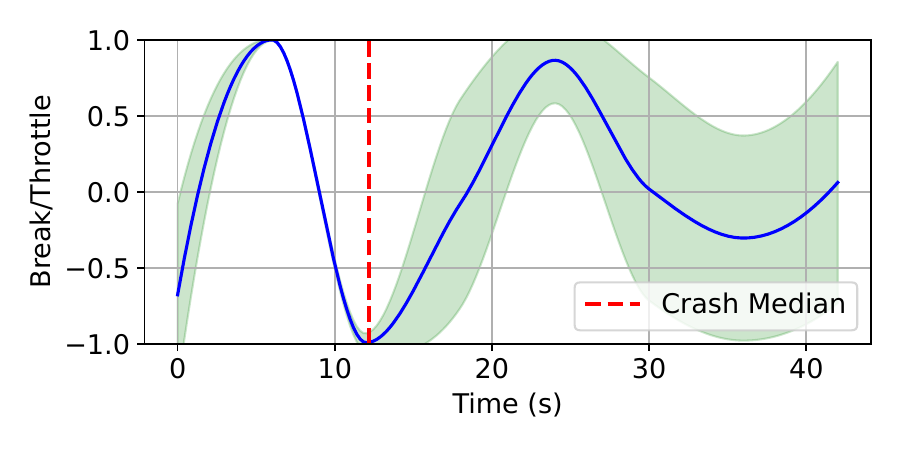}
    \caption{Cluster 1 mean and standard deviation Bayesian Optimization}\label{fig:stddev-bo-1-rest}
\end{subfigure}
\begin{subfigure}[b]{.4\linewidth}
    \includegraphics[width=\linewidth]{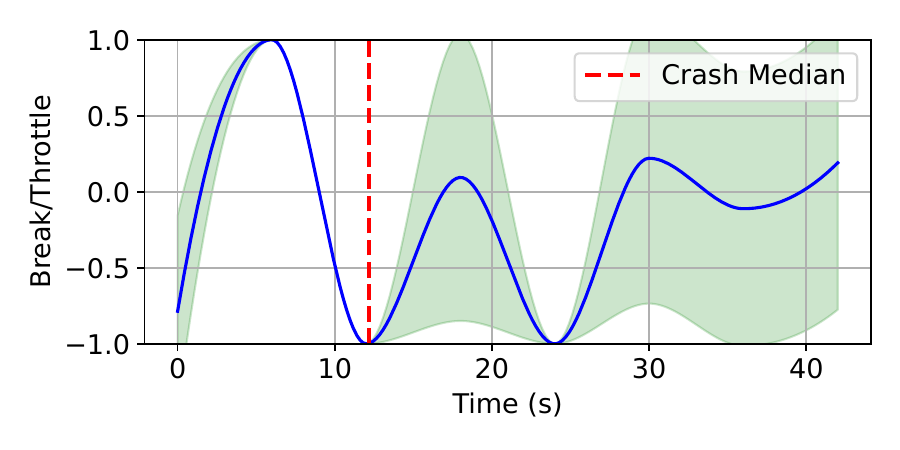}
    \caption{Cluster 2 mean and standard deviation Bayesian Optimization}\label{fig:stddev-bo-2-rest}
\end{subfigure}

\caption{Successful attack maneuvers for Bayesian Optimization with speed 25 and setpoint distance 7 (transient period) }
\label{fig:stddev-bo-rest}\vspace{-2ex}
\end{figure}


To ensure this maneuver was the most significant, we further describe the behavior of each cluster in Fig.~\ref{fig:stddev-bo-rest}. In it we depict the mean and standard deviation of the attacks in each cluster with the overall median of all successful attacks in those particular conditions. Fig~\ref{fig:stddev-bo-1-rest} and Fig~\ref{fig:stddev-bo-2-rest} show almost the exact same values for the first three parameters, with the median indicating that most of the attacks happen within the range of 0 to 12 seconds. Furthermore, inspecting the cluster with the most attacks, Cluster 1, we can see a pattern emerging with another peak of acceleration followed by a brake with a lower standard deviation than in Cluster 2. This leads us to conclude that during a transient period, the attacker prefers a maneuver of max brake to max acceleration.

\begin{figure}[htb]\vspace{-2ex}
\centering
\begin{subfigure}[b]{.4\linewidth}
    \includegraphics[width=\linewidth]{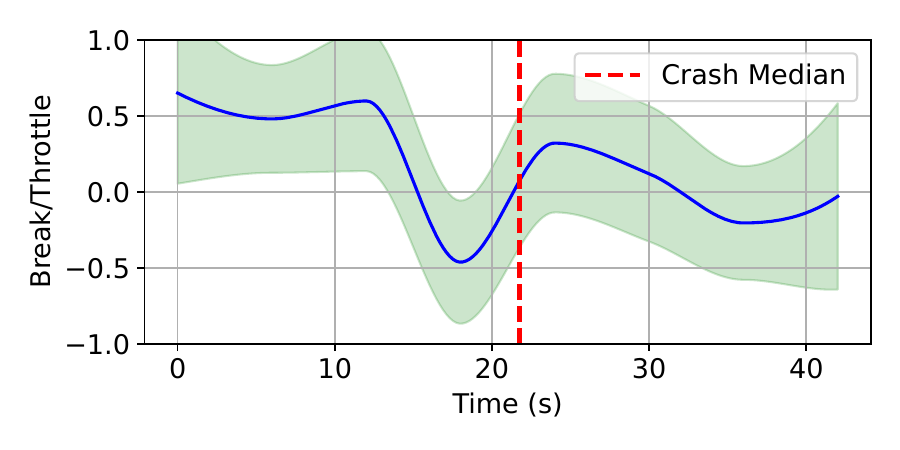}
    \caption{Cluster 1 mean and standard deviation Cross-Entropy}\label{fig:stddev-ce-1-rest}
\end{subfigure}
\begin{subfigure}[b]{.4\linewidth}
    \includegraphics[width=\linewidth]{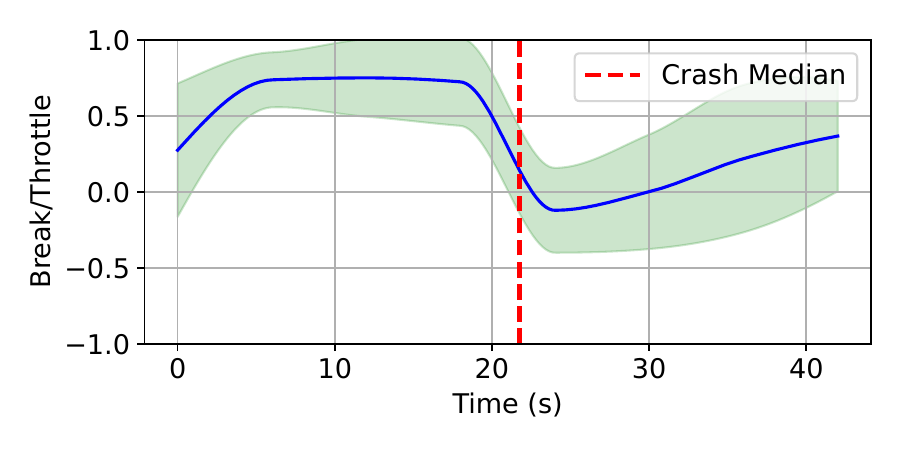}
    \caption{Cluster 2 mean and standard deviation Cross-Entropy}\label{fig:stddev-ce-2-rest}
\end{subfigure}

\caption{Successful attack maneuvers for Bayesian Optimization with speed 25 and setpoint distance 7 (transient period) }
\label{fig:stddev-ce-rest}\vspace{-2ex}
\end{figure}

\begin{figure}[h!]\vspace{-2ex}
\centering
\begin{subfigure}[b]{.4\linewidth}
    \includegraphics[width=\linewidth]{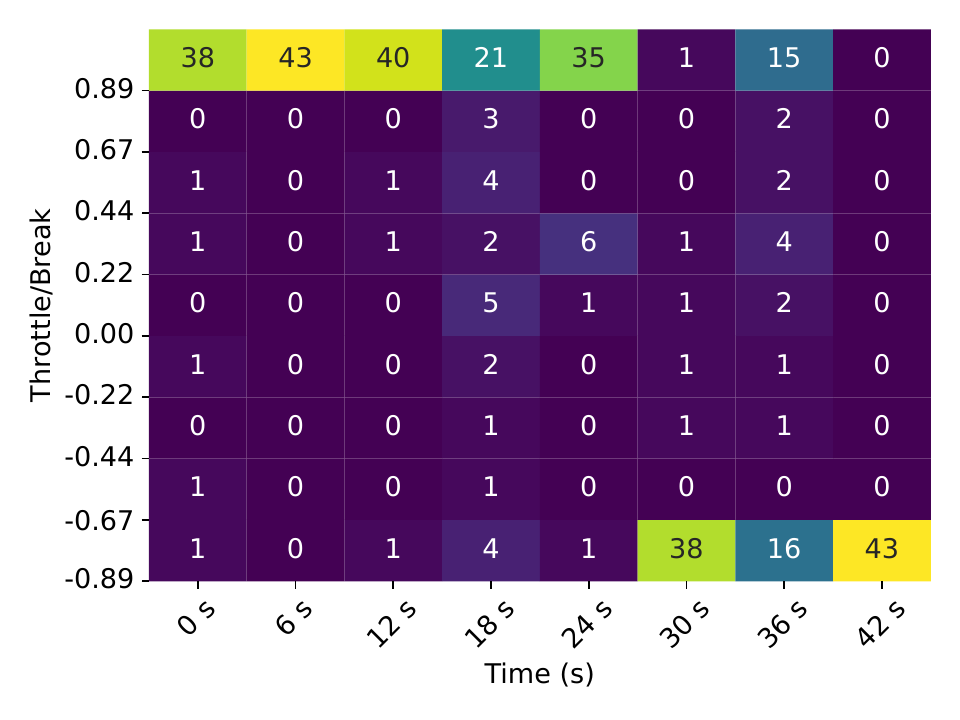}
    \caption{Cluster 1 Bayesian Optimization}\label{fig:clusterCE_1_rest}
\end{subfigure}
\begin{subfigure}[b]{.4\linewidth}
    \includegraphics[width=\linewidth]{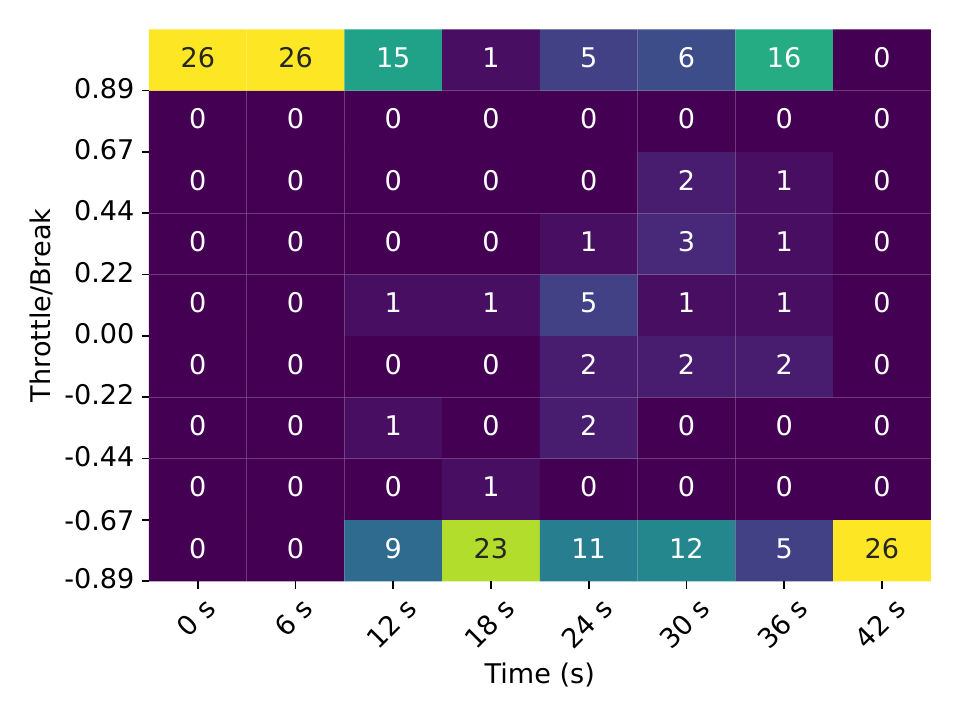}
    \caption{Cluster 2 Bayesian Optimization}\label{fig:clusterCE_2_rest}
\end{subfigure}
\begin{subfigure}[b]{.4\linewidth}
    \includegraphics[width=\linewidth]{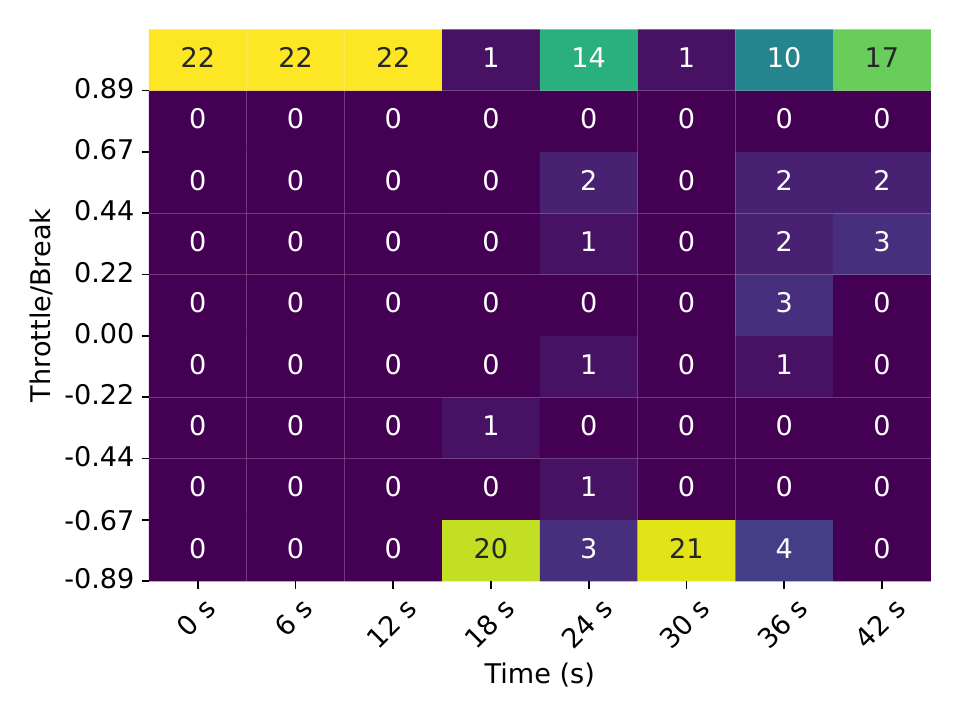}
    \caption{Cluster 3 Bayesian Optimization}\label{fig:clusterCE_2_rest}
\end{subfigure}

\caption{Values of variables in each successful attack in each cluster for Bayesian Optimization with speed 20 and setpoint distance 5 (after 15 s) }

\label{fig:clusterbo_15}\vspace{-2ex}
\end{figure}

For comparison, we analyze the CE clusters, these were considerably more ineffective in finding successful attacks. In Fig.~\ref{fig:stddev-ce-rest}  we notice that cluster two has a more defined pattern with less variation than cluster one and aligns its brake actions with the median of the attacks. The alignment allows us to conclude that CE mainly focuses on finding one specific action point to brake, avoiding BO's oscillation pattern, which can explain why is less successful when finding attacks.

\subsection{ Attacks after steady state period}
We perform the same analysis for the attacks that start after the vehicles reach a steady state speed and correspondingly perform clustering, aiming to discover patterns selected by the attacker. In this case, we divide BO into three clusters by combining elbow rule and DBSCAN measurements. These clusters confirm the tendency of this sampler to focus on the extreme values to get the attacks, and, contrary to the previous scenario, the attacker accelerates on the first three action points instead of braking. 

\noindent \textbf{Attacker maneuver:}Comparing clusters in Fig. \ref{fig:clusterBO_rest} with the ones in Fig. \ref{fig:clusterbo_15} shows an important difference. While the transient attack focuses on instant braking initially, in this case, after stabilization, the attacker applies maximum acceleration most of the time for up to 20 seconds and aims to increase the space as much as possible before performing a hard stop. Afterward, the attacker mainly accelerates at 36 seconds, maximizing distance, and brakes again, trying to generate an oscillation. However, cluster 2's strategy is much more diverse after the action point in 18 seconds, which indicates that the attacks in this cluster do not find a clear attack signal as easily. We can extract an attack maneuver that emerges from these clusters, particularly in the first 3 timesteps that we illustrate in Alg.~\ref{algo:man2} 

\begin{algorithm}
    \caption{Steady State Maneuver}
    \begin{algorithmic}

        \State \textbf{Apply  Full Throttle} at $t \approx 0 $
        \State \textbf{Apply  Max Brake} at $t\approx 30$
        \State \textbf{Apply  Full Throttle} at $t\approx 36$
        \If{no crash}
            \State \textbf{Apply  Max Brake} at $t\approx 42$
        \EndIf

    \end{algorithmic}
    \label{algo:man2}
\end{algorithm}

\begin{figure}[htb]\vspace{-2ex}
\centering
\begin{subfigure}[b]{.4\linewidth}
    \includegraphics[width=\linewidth]{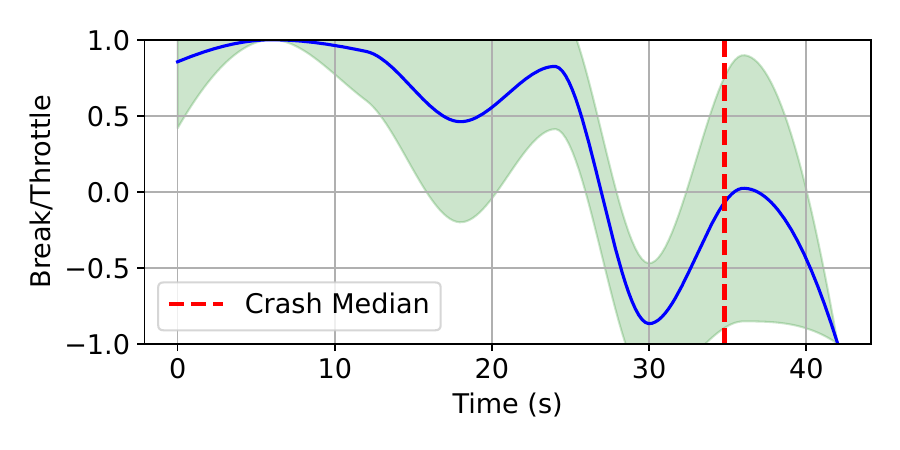}
    \caption{Cluster 1 mean and standard deviation Bayesian Optimization}\label{fig:stddev-bo-1-15}
\end{subfigure}
\begin{subfigure}[b]{.4\linewidth}
    \includegraphics[width=\linewidth]{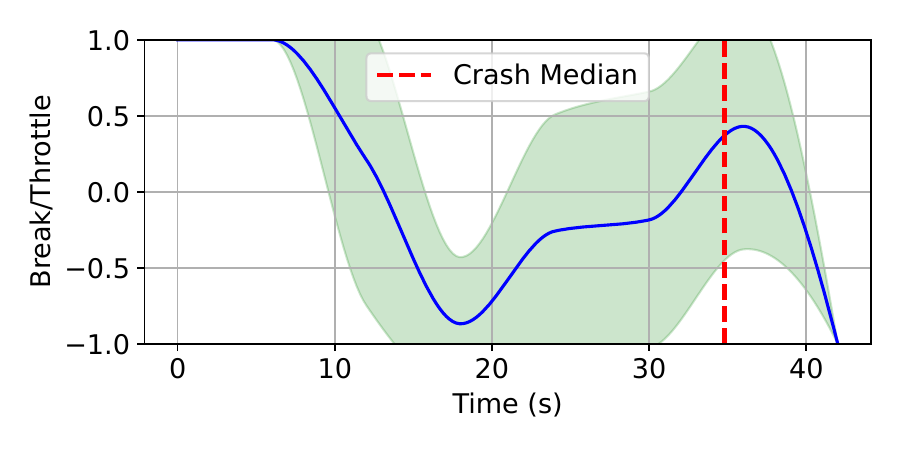}
    \caption{Cluster 2 mean and standard deviation Bayesian Optimization}\label{fig:stddev-bo-2-15}
\end{subfigure}
\begin{subfigure}[b]{.4\linewidth}
    \includegraphics[width=\linewidth]{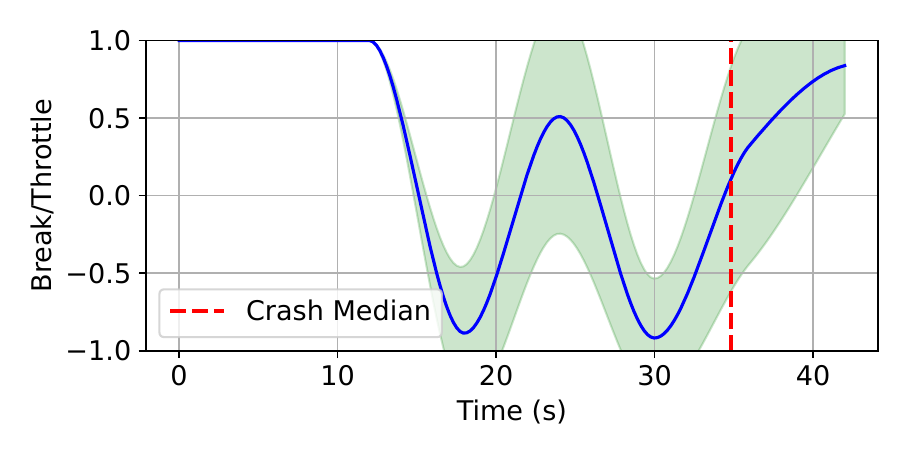}
    \caption{Cluster 3 mean and standard deviation Bayesian Optimization}\label{fig:stddev-bo-3-15}
\end{subfigure}

\caption{Successful attack maneuvers for Bayesian Optimization with speed 20 and setpoint distance 5 (after steady state period) }
\label{fig:stddev-bo-15}\vspace{-2ex}
\end{figure}

Looking further into the clusters in Fig. \ref{fig:stddev-bo-15}, the median line reveals that the most significant action points are 36 and 42 seconds, as half of the attacks occur here. This could indicate that the final braking action taken by the biggest clusters, one and two, in their last action points is the most significant. In contrast, the most consistent oscillation pattern in cluster 3 (Fig. \ref{fig:stddev-bo-3-15}) could result from most attacks before the median line.





\begin{figure}[htb]\vspace{-2ex}
\centering
\begin{subfigure}[b]{.4\linewidth}
    \includegraphics[width=\linewidth]{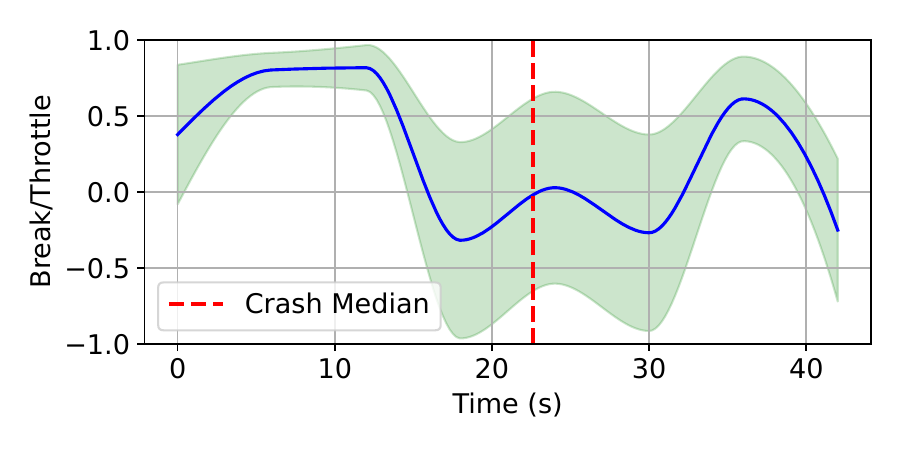}
    \caption{Cluster 1 mean and standard deviation Cross-Entropy}\label{fig:stddev-ce-1-15}
\end{subfigure}
\begin{subfigure}[b]{.4\linewidth}
    \includegraphics[width=\linewidth]{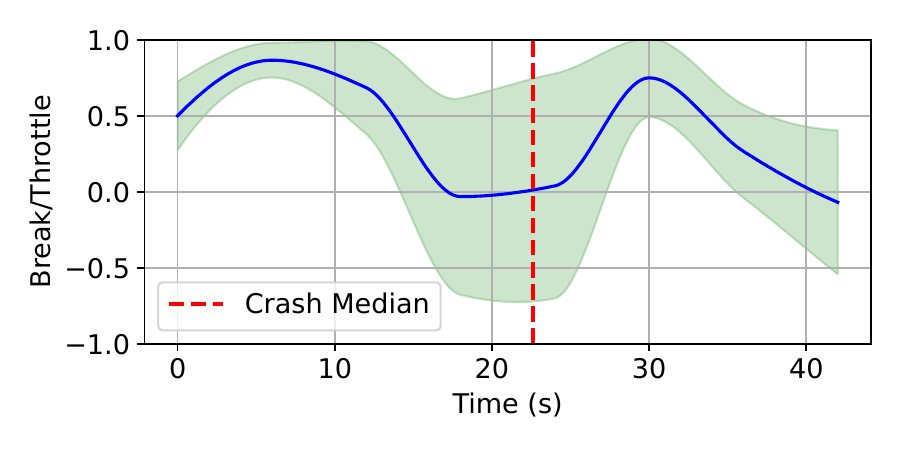}
    \caption{Cluster 2 mean and standard deviation Cross-Entropy}\label{fig:stddev-ce-2-15}
\end{subfigure}

\caption{Successful attack maneuvers for Cross-Entropy with speed 20 and setpoint distance 7 (after steady state period) }
\label{fig:stddev-ce-15}\vspace{-2ex}
\end{figure}

In comparison, CE has considerably less success in this area. However, there is a similarity at the beginning of the attacks in Fig. \ref{fig:stddev-ce-15} with an acceleration and then a steady brake that, contrary to BO, has a more measured approach. In this case, the median does not provide much insight as its position closely aligns with the center of the attack, making it challenging to infer which braking action effectively caused the attack. 
\section{Conclusion}
Our results can help guide vehicle designers in maintaining a minimum distance between vehicles. We also highlight the danger that the possibility of speed differences above $15 km/h$ can have on a potential crash.
Likewise, our results show how an attacker can manufacture different patterns with diverse acceleration and braking actions to successfully cause a crash. This indicates that a clever attacker can develop systematic patterns that significantly increase the chance of a crash and show that extreme changes in acceleration can cause many crashes. ACC controllers are also vulnerable to more complex attacks with more variations. 

While malicious drivers can also cause crashes with human drivers, there are two main differences in testing these attacks with specific autonomous vehicles: (1) If the attacker finds an attack, it can be exploited and reproduced with predictable consequences in a large-scale. (2) The attacker has the confidence that the tested autonomous vehicle will not crash the malicious driver. In contrast, attacking human drivers will have more unpredictable consequences and might end in the attacker's vehicle being rear-ended more often than not. 

The experiments show that BO is consistently more effective at discovering attacks than CE. We can attribute this to the way BO works, prioritizing the best point based on prior knowledge, and is generally more effective when optimizing a lower amount of variables. CE, on the other hand, works by sampling over a distribution and requires a more significant amount of iterations to find the best solution. As our experiments are limited to 100 samples per configuration, it is not enough for CE to get the best solution. 

However, there are disparities in how BO and CE perform in our experiments. In our parametric attacks, the difference in the number of attacks between CE and BO is significantly lower than in non-parametric attacks. This is driven by the fact that our parametric attacks restrict the search space and have fewer variables to consider which benefits CE.

It is also important to highlight our newly discovered attack maneuvers; we find that overall attacks after the steady state are more effective, with the attacker favoring a pattern of accelerating to gain distance and then braking and, when failing, trying again. This pattern is followed mainly by Cluster 1 in Fig. \ref{fig:clusterCE_1_rest} with the most successful attacks in its configuration and most often crashing the other vehicles in points between 36 and 42 on its second try. 

To mitigate these vulnerabilities ACCs could implement setpoint distances that show resilience in our experiments. In Fig. \ref{fig:headmaps}, the attacker is least successful in cases where the setpoint distance is at least 11 meters between vehicles in both steady and transient states. This implementation could considerably reduce the effects of an attack, particularly at slower target speeds of 20 m/s. Moreover, ACC controllers could adopt strategies such as \cite{doi:10.1243/095440704772913918} that can handle different distances at different speeds. 

In future work, we plan to consider more sophisticated malicious drivers that swerve in and out of traffic, as well as more sophisticated autonomous vehicle stacks, such as Apollo and Autoware.

\subsubsection{Acknowledgments}
This material is based upon work supported in part by the Air Force Office of Scientific Research under award number FA9550-24-1-0015, by the Google-CAHSI IRP program, and by the National Center for Transportation Cybersecurity and Resiliency (TraCR) (a U.S. Department of Transportation National University Transportation Center--USDOT Grant \#69A3552344812)). 

\bibliographystyle{splncs04}
\bibliography{bibliography}

\begin{thebibliography}{10}
\providecommand{\url}[1]{\texttt{#1}}
\providecommand{\urlprefix}{URL }
\providecommand{\doi}[1]{https://doi.org/#1}

\bibitem{akazaki2018falsification}
Akazaki, T., Liu, S., Yamagata, Y., Duan, Y., Hao, J.: Falsification of cyber-physical systems using deep reinforcement learning. In: Formal Methods: 22nd International Symposium, FM 2018, Held as Part of the Federated Logic Conference, FloC 2018, Oxford, UK, July 15-17, 2018, Proceedings 22. pp. 456--465. Springer (2018)

\bibitem{blasch2021certifiable}
Blasch, E., Bin, J., Liu, Z.: Certifiable artificial intelligence through data fusion. arXiv preprint arXiv:2111.02001  (2021)

\bibitem{clarke2018handbook}
Clarke, E.M., Henzinger, T.A., Veith, H., Bloem, R., et~al.: Handbook of model checking, vol.~10. Springer (2018)

\bibitem{corso2021survey}
Corso, A., Moss, R., Koren, M., Lee, R., Kochenderfer, M.: A survey of algorithms for black-box safety validation of cyber-physical systems. Journal of Artificial Intelligence Research  \textbf{72},  377--428 (2021)

\bibitem{daganzo1997fundamentals}
Daganzo, C.F.: Fundamentals of transportation and traffic operations. Emerald Group Publishing Limited (1997)

\bibitem{darema2023handbook}
Darema, F., Blasch, E., Ravela, S., Aved, A.J.: Handbook of Dynamic Data Driven Applications Systems, vol.~2. Springer (2023)

\bibitem{de2005tutorial}
De~Boer, P.T., Kroese, D.P., Mannor, S., Rubinstein, R.Y.: A tutorial on the cross-entropy method. Annals of operations research  \textbf{134},  19--67 (2005)

\bibitem{deshmukh2019formal}
Deshmukh, J.V., Sankaranarayanan, S.: Formal techniques for verification and testing of cyber-physical systems. Design Automation of Cyber-Physical Systems pp. 69--105 (2019)

\bibitem{dosovitskiy2017carla}
Dosovitskiy, A., Ros, G., Codevilla, F., Lopez, A., Koltun, V.: Carla: An open urban driving simulator (2017)

\bibitem{duggirala2015c2e2}
Duggirala, P.S., Mitra, S., Viswanathan, M., Potok, M.: C2e2: A verification tool for stateflow models. In: Tools and Algorithms for the Construction and Analysis of Systems: 21st International Conference, TACAS 2015, Held as Part of the European Joint Conferences on Theory and Practice of Software, ETAPS 2015, London, UK, April 11-18, 2015, Proceedings 21. pp. 68--82. Springer (2015)

\bibitem{elvik2009power}
Elvik, R.: The Power Model of the relationship between speed and road safety: update and new analyses. No. 1034/2009 (2009)

\bibitem{10.5555/3001460.3001507}
Ester, M., Kriegel, H.P., Sander, J., Xu, X.: A density-based algorithm for discovering clusters in large spatial databases with noise. In: Proceedings of the Second International Conference on Knowledge Discovery and Data Mining. p. 226–231. KDD'96, AAAI Press (1996)

\bibitem{fainekos2006robustness}
Fainekos, G.E., Pappas, G.J.: Robustness of temporal logic specifications. In: International Workshop on Formal Approaches to Software Testing. pp. 178--192. Springer (2006)

\bibitem{fainekos2009robustness}
Fainekos, G.E., Pappas, G.J.: Robustness of temporal logic specifications for continuous-time signals. Theoretical Computer Science  \textbf{410}(42),  4262--4291 (2009)

\bibitem{fremont2020formal}
Fremont, D.J., Kim, E., Pant, Y.V., Seshia, S.A., Acharya, A., Bruso, X., Wells, P., Lemke, S., Lu, Q., Mehta, S.: Formal scenario-based testing of autonomous vehicles: From simulation to the real world. In: 2020 IEEE 23rd International Conference on Intelligent Transportation Systems (ITSC). pp.~1--8. IEEE (2020)

\bibitem{f1aac6cb-75ac-3b50-89ba-5fa72fe91ebe}
Fritsch, F.N., Carlson, R.E.: Monotone piecewise cubic interpolation. SIAM Journal on Numerical Analysis  \textbf{17}(2),  238--246 (1980), \url{http://www.jstor.org/stable/2156610}

\bibitem{haspalamutgil2017adaptive}
Haspalamutg{\i}l, K., Adali, E.: Adaptive switching method for adaptive cruise control. In: 2017 21st International Conference on System Theory, Control and Computing (ICSTCC). pp. 140--145. IEEE (2017)

\bibitem{hernandez2024d4}
Hernandez, C., Ortiz~Barbosa, D.E., Lei, Z., Burbano, L., Park, Y., Ukkusuri, S.V., Cardenas, A.: D4: Dynamic data-driven discovery of adversarial vehicle maneuvers. In: Proceedings of the DDDAS 2024. New Brunswick, NJ (November 2024)

\bibitem{ivanov2019verisig}
Ivanov, R., Weimer, J., Alur, R., Pappas, G.J., Lee, I.: Verisig: verifying safety properties of hybrid systems with neural network controllers. In: Proceedings of the 22nd ACM International Conference on Hybrid Systems: Computation and Control. pp. 169--178 (2019)

\bibitem{kapinski2016simulation}
Kapinski, J., Deshmukh, J.V., Jin, X., Ito, H., Butts, K.: Simulation-based approaches for verification of embedded control systems: An overview of traditional and advanced modeling, testing, and verification techniques. IEEE Control Systems Magazine  \textbf{36}(6),  45--64 (2016)

\bibitem{kapinski2014simulation}
Kapinski, J., Deshmukh, J.V., Sankaranarayanan, S., Arechiga, N.: Simulation-guided lyapunov analysis for hybrid dynamical systems. In: Proceedings of the 17th international conference on Hybrid systems: computation and control. pp. 133--142 (2014)

\bibitem{10.1145/2562059.2562139}
Kapinski, J., Deshmukh, J.V., Sankaranarayanan, S., Arechiga, N.: Simulation-guided lyapunov analysis for hybrid dynamical systems. In: Proceedings of the 17th International Conference on Hybrid Systems: Computation and Control. p. 133–142. HSCC '14, Association for Computing Machinery, New York, NY, USA (2014). \doi{10.1145/2562059.2562139}, \url{https://doi.org/10.1145/2562059.2562139}

\bibitem{Khalil:1173048}
Khalil, H.K.: {Nonlinear systems; 3rd ed.} Prentice-Hall, Upper Saddle River, NJ (2002), \url{https://cds.cern.ch/record/1173048}, the book can be consulted by contacting: PH-AID: Wallet, Lionel

\bibitem{koren2018adaptive}
Koren, M., Alsaif, S., Lee, R., Kochenderfer, M.J.: Adaptive stress testing for autonomous vehicles. In: 2018 IEEE Intelligent Vehicles Symposium (IV). pp.~1--7. IEEE (2018)

\bibitem{1056489}
Lloyd, S.: Least squares quantization in pcm. IEEE Transactions on Information Theory  \textbf{28}(2),  129--137 (1982). \doi{10.1109/TIT.1982.1056489}

\bibitem{mockus1989bayesian}
Mockus, J.: Bayesian Approach to Global Optimization: Theory and Applications. Kluwer Academic (1989)

\bibitem{paul2023formal}
Paul, S., Cruz, E., Dutta, A., Bhaumik, A., Blasch, E., Agha, G., Patterson, S., Kopsaftopoulos, F., Varela, C.: Formal verification of safety-critical aerospace systems. IEEE Aerospace and Electronic Systems Magazine  \textbf{38}(5),  72--88 (2023)

\bibitem{romeijn1994simulated}
Romeijn, H.E., Smith, R.L.: Simulated annealing for constrained global optimization. Journal of Global Optimization  \textbf{5},  101--126 (1994)

\bibitem{salgado2022fuzzing}
Salgado, I.F., Quijano, N., Fremont, D.J., Cardenas, A.A.: Fuzzing malicious driving behavior to find vulnerabilities in collision avoidance systems. In: 2022 IEEE European Symposium on Security and Privacy Workshops (EuroS\&PW). pp. 368--375. IEEE (2022)

\bibitem{cbc87a20-8264-374b-b836-e5a5f6e270e4}
Sanfelice, R.G.: Hybrid Feedback Control. Princeton University Press (2021), \url{http://www.jstor.org/stable/j.ctv131btfx}

\bibitem{tran2020nnv}
Tran, H.D., Yang, X., Manzanas~Lopez, D., Musau, P., Nguyen, L.V., Xiang, W., Bak, S., Johnson, T.T.: Nnv: the neural network verification tool for deep neural networks and learning-enabled cyber-physical systems. In: International Conference on Computer Aided Verification. pp. 3--17. Springer (2020)

\bibitem{doi:10.1243/095440704772913918}
Wang, J., Rajamani, R.: The impact of adaptive cruise control systems on highway safety and traffic flow. Proceedings of the Institution of Mechanical Engineers, Part D  \textbf{218}(2),  111--130 (2004). \doi{10.1243/095440704772913918}, \url{https://doi.org/10.1243/095440704772913918}

\bibitem{wang2020data}
Wang, X., Shen, S., Bezzina, D., Sayer, J.R., Liu, H.X., Feng, Y.: Data infrastructure for connected vehicle applications. Transportation Research Record  \textbf{2674}(5),  85--96 (2020)

\bibitem{zhenhai2016multi}
Zhenhai, G., Jun, W., Hongyu, H., Wei, Y., Dazhi, W., Lin, W.: Multi-argument control mode switching strategy for adaptive cruise control system. Procedia engineering  \textbf{137},  581--589 (2016)

\end{thebibliography}

\end{document}